\documentclass[a4paper,11pt]{article}
\usepackage{blindtext}
\usepackage{geometry}
\usepackage{multirow}
\usepackage[utf8]{inputenc}
\usepackage{amsmath}
\usepackage{amsfonts}
\usepackage{enumitem}
\usepackage{longtable}
\usepackage{array}
\usepackage{subcaption}
\usepackage{graphicx}
\usepackage{ulem}

\usepackage{float}  
\newgeometry{vmargin={30mm}, hmargin={20mm,20mm}} 
\usepackage{graphicx} 
\usepackage[style=numeric-comp, sorting=none, backend=biber]{biblatex}
\usepackage{amssymb}
\usepackage{hyperref}
\usepackage{authblk}  
\usepackage{tikz-feynman}
\tikzfeynmanset{compat=1.1.0} 
\usepackage{amsmath}

\title{\textbf{The Alternative Left–Right Scenario: Unitarity, Vacuum Stability 
and RG Evolution}}
\author[1]{\textbf{Hrishikesh Deka}\thanks{\texttt{hrishikesh.deka@iitg.ac.in}}}
\author[2]{\textbf{Avnish}\thanks{\texttt{avnishy@uohyd.ac.in}}}
\author[3]{\textbf{Sumit K. Garg}\thanks{\texttt{sumit.kumar@manipal.edu}}}
\author[4]{\textbf{Poulose Poulose}\thanks{\texttt{poulose@sju.edu.in}}}

\affil[1]{Department of Physics, Indian Institute of Technology Guwahati, North Guwahati, 781039, India.}
\affil[2]{School of Physics, University of Hyderabad, Hyderabad - 500046, India.}
\affil[3]{Manipal Centre for Natural Sciences, Manipal Academy of Higher Education, Dr.T.M.A. Pai Planetarium Building, Manipal-576104, Karnataka, India}
\affil[4]{Department of Physics, St. Joseph's University, Bangalore 560 027, India}
\begin{document}
 \date{}
\maketitle

\begin{abstract}
We study the theoretical constraints on the scalar sector of the Alternative Left-Right 
Model (ALRM), an $E_6$-motivated extension of the Standard Model based on the gauge group 
$\mathrm{SU}(3)_c \otimes \mathrm{SU}(2)_L \otimes \mathrm{SU}(2)_{R'} \otimes 
\mathrm{U}(1)_{B-L}$, supplemented by a global $\mathrm{U}(1)_S$ symmetry. We derive the complete set of tree-level perturbative unitarity constraints on the model, resulting in 14 independent conditions on the quartic scalar couplings. When combined with the boundedness-from-below conditions and the requirement of positive-definite scalar mass-squared eigenvalues, these constraints are found to be complementary, with their simultaneous imposition yielding significantly more stringent restrictions on the parameter space than either set alone. We then perform a one loop renormalization group analysis, evolving the model parameters from the electroweak scale up to a high energy cut-off scale, and requiring that the vacuum stability, the unitarity, and the perturbativity conditions are preserved throughout. The renormalisation group evolution is found to restrict the allowed parameter space considerably beyond the tree-level bounds, with the constraints on the quartic couplings becoming more stringent as the cut off scale is raised. Consequently, the physical scalar masses in the model acquire upper bounds. For the right-hand symmetry breaking scale, $v_R = 10$~TeV and requiring theoretical consistency up to $10^{16}$~GeV, we obtain $m_{H_1^\pm} \lesssim 6.5$~TeV, $m_{H_2^\pm} \lesssim 1.5$~TeV, and $m_{H_1^0} \simeq m_{A_1} \lesssim 1.3$~TeV, with all bounds scaling with $v_R$. These findings offer a predictive and falsifiable framework for searches of the extended Higgs sector of the ALRM at the current and future collider experiments.
\end{abstract}

\section{Introduction}
While being remarkably successful in describing the dynamics of elementary particles at the electroweak scale, the Standard Model (SM) of particle physics fails to account for several critical experimental and theoretical shortcomings. The observation of nonzero neutrino masses, the overwhelming evidence for dark matter, and the baryon asymmetry of the universe (BAU) are among the most compelling experimental shortcomings that point towards the incompleteness of the SM. Whereas, on the theoretical side, concerns on the stability of the electroweak vacuum and the gauge hierarchy problem represent the issues of fundamental in nature that the Standard Model has remained unable to address adequately yet. Among several attempts to resolve these difficulties, phenomenological models hold a central place. A strongly prevalent notion among particle physicists is that larger symmetries unify the dynamics at some higher energy scale, and break down naturally to the SM gauge group at the electroweak scale. Along this line of thought, several grand unified symmetries have been proposed and studied. Of particular interest in this work is the exceptional group $E_6$, whose symmetry breaking proceeds through multiple steps. It was observed decades ago that this breaking chain admits an intermediate stage where the dynamics is left-right symmetric. This allowed a natural explanation for the otherwise puzzling observation of maximal parity violation in weak interactions, which is introduced as an input within the SM. As a bonus, this scenario also provides a natural mechanism for generating nonzero neutrino masses via seesaw mechanism. 

The conventional Left-Right Symmetric Model (LRSM) features an additional $SU(2)_R$ gauge group, under which the right-handed up-type quarks and the right-handed down-type quarks pair up as doublets. Similarly, the right-handed charged lepton is partnered with a right-handed neutrino. All these fields, including the right-handed neutrinos, are accommodated within the 27-plet of $E_6$. The LRSM, however, suffers severely from experimental constraints in the flavour sector. In particular, it predicts large flavour-changing neutral current (FCNC) interactions, whereas the experimental bounds on such interactions are extremely stringent. Moreover, the LRSM offers no viable dark matter candidate and struggles to accommodate the dynamics necessary to explain the baryon asymmetry of the universe (BAU). However, an alternative scenario, named Alternative Left-Right Model (ALRM) emerges when the members of the 
27-plet of $E_6$ are arranged into different multiplets of $SU(2)\times SU(2)$ compared to the conventional LRSM. In LRSM these gauge groups are interpreted as $SU(2)_L\times SU(2)_R$. While in the ALRM, the right-handed down-type quark $d_R$ no longer partners with the right-handed up-type quark under right-handed gague group. Instead, a new exotic quark $d^{\prime}_R$ assumes this role, leaving $d_R$ as a singlet under both $SU(2)_L$ and $SU(2)_R$. To distinguish this behavior, we denote the gauge group of ALRM as $SU(2)_L\times SU(2)_{R'}$. An analogous re-arrangement occurs in the leptonic sector, where the right-handed neutrino is treated as a singlet under all gauge symmetries. 
As discussed in detail in the subsequent sections, this re-arrangement permits the introduction of a global symmetry that forbids the otherwise dangerous tree-level FCNC interactions. Remarkably, this same symmetry provides a dark matter candidate, stabilized by the remnant $R$-parity surviving in the symmetry-broken phase.

The scalar sector of ALRM consists of one Higgs bi-doublet $\Phi$ and two Higgs doublets denoted as $\chi_{L,R}$ (one each under $SU(2)_L$ and $SU(2)_{R'}$, respectively). The richer scalar field content of this scenario admits a stable vacuum configuration more readily than the SM. A systematic study of the scalar potential was carried out in Ref.~\cite{Frank:2021ekj}, where the parameters of the tree-level potential were constrained by requiring boundedness from below (BFB). Expressing the physical scalar masses in terms of these parameters and the vacuum expectation values, the additional requirement of positive-definite mass-squared eigenvalues imposes a further set of constraints, collectively restricting the viable parameter space of the model. Similar analyses of stable vacuum structure in multi-scalar potentials~\cite{Kannike:2012pe, Kannike:2016fmd, Sanchez-Vega:2018qje, Chakrabortty:2013zja, Chakrabortty:2013mha, Kim:1981xu, Kannike:2021fth} and in the conventional Left-Right Symmetric Model (LRSM)~\cite{BhupalDev:2018xya, Chauhan:2019fji, Branco:1985ng, Basecq:1985sx, Mohapatra:2014qva, Brahmachari:1994ts} have been extensively explored in the literature. Another key theoretical constraint on the Higgs sector arises from perturbative unitarity~\cite{Gell-Mann:1969cuq,Weinberg:1971fb,Lee:1977yc,
Lee:1977eg}. This requirement, coming from the unitarity of the S-matrix, implies that partial-wave scattering amplitudes must remain bounded in magnitude. These constraints are being used to put bounds~\cite{Lee:1977yc,
Lee:1977eg} on the mass of Higgs boson. The perturbative unitarity constraints have not yet been examined in the context of the alternative Left-Right Model (ALRM), and we aim to address them in this work.

Furthermore, the previous studies in the ALRM scenario have demonstrated the model’s potential implications for neutrinoless double beta decay and leptogenesis in Ref.~\cite{Frank:2020odd}. It have also been shown that exotic fermions and Higgs bosons can indirectly affect rare top decays ~\cite{Frank:2023fkc}, as well as contribute to rare lepton-flavour-violating (LFV) processes and the anomalous magnetic moment of the muon, $a_{\mu}$~\cite{Frank:2024bss, Frank:2025pvc}. Building on this, in the present work, we impose all tree level theoretical constraints, namely the potential being bounded from below, perturbative unitarity constraints, and the perturbativity limits of the couplings, and check their stability under one loop renormalization group evolution. By tracking the running of the couplings and mass parameters, we identify the energy scales up to which these conditions continue to hold. Bringing these elements together in a single framework, we then determine the corresponding constraints on the parameter space of the ALRM.


The organization of the paper is as follows: in Sec.~\ref{sec:Model}, we present briefly the relevant details of the model framework, followed by a discussion of the theoretical constraints in Sec.~\ref{sec:theoretical}. This section also includes a numerical analysis identifying the region of parameter space compatible with the BFB and perturbative unitarity constraints. In Sec.~\ref{sec:RGE}, we study the renormalization group evolution (RGE) of the couplings in detail and numerically analyse its effect on these constraints. Finally, we conclude our findings in Sec.~\ref{sec:conclusion}.


\section{Model framework}
\label{sec:Model}
In this section, we briefly review the alternative Left-Right Model (ALRM) framework and refer the interested reader to Ref.~\cite{Frank:2019nid} for a more detailed discussion. The ALRM framework is based on the gauge group \(
SU(3)_c \otimes SU(2)_L \otimes SU(2)_{R'} \otimes U(1)_{B-L}\) supplemented by an additional global symmetry ${ U}(1)_{S}$. 
The complete scalar and fermionic field content of the model, along with the corresponding gauge quantum numbers, is given in Table~\ref{tab:my_label}. The electric charges of the particle fields can be derived from the generalised Gell-Mann-Nishijima relation $Q= T_{3R'}+ T_{3L} + Y_{B-L}$, where $Y_{B-L}$ is the ${ U}(1)_{\rm B-L}$ charge, and $T_{3L}$ and $T_{3R'}$ are 
the third-component isospin quantum numbers under  ${  SU}(2)_{\rm L}$ and ${ SU}(2)_{\rm R'}$, respectively.

\begin{table}[h]
 \centering
 \begin{tabular}{ c | c | c | c | c | c||c |c} 
		\hline \hline
		Particles  &${ SU}(3)_{ C}$  &${ SU}(2)_{ L}$  &${ SU}(2)_{{\rm R'}}$  &${ U}(1)_{{B-L}}$   &${
 U}(1)_{ S}$ 
        &$L$&$R$\\
       \hline
		
       $Q_L$ = $\begin{pmatrix}
		u_L \\ 
		d_L 
		\end{pmatrix}$ &3 &2 &1 &$\frac{1}{6}$ &0&0&+ \\ [5mm]
        
       $Q_R$=$\begin{pmatrix}
		u_R \\ 
		d^{\prime}_{R}
		\end{pmatrix}$ &3 &1 &2 &$\frac{1}{6}$ & - $\frac{1}{2}$&$\begin{pmatrix}
		0 \\ 
		-1
		\end{pmatrix}$&$\begin{pmatrix}
		+ \\ 
		-
		\end{pmatrix}$\\[5mm] 
       
       $d^{\prime}_{L}$    &3 &1 &1 & - $\frac{1}{3}$ &$-1$&$-1$&$-$  \\ 
       
       $d_{R}$        &3 &1 &1 &- $\frac{1}{3}$ &0&0&+ \\  
       \hline
        
       $L_L$ = $\begin{pmatrix}
		\nu_L \\ 
		e_L 
		\end{pmatrix}$  &1 &2 &1 &-$\frac{1}{2}$ & 1&1&+\\  [5mm]  
       
       $L_R$ = $\begin{pmatrix}   
		n_R \\ 
		e_R 
		\end{pmatrix}$  &1 &1 &2 &-$\frac{1}{2}$ &  $\frac{3}{2}$&$\begin{pmatrix}   
		2 \\ 
		1 
		\end{pmatrix}$ &$\begin{pmatrix}   
		- \\ 
		+ 
		\end{pmatrix}$ \\
        
       $n_L$           &1 &1 &1 &0 &2&2&$-$\\
        
       $\nu_R$         &1 &1 &1 &0 &1&1&+\\
		\hline  
       
       $\Phi$ = $\begin{pmatrix}
	    \phi_1^{0}  & \phi_1^{+} \\ 
	    \phi_2^{-}  & \phi_2^{0}
	    \end{pmatrix}$ &1 &2 &$2^*$ &0 & -$\frac{1}{2}$
        &$\begin{pmatrix}
	    -1  & 0 \\ 
	    -1 & 0
	    \end{pmatrix}$&	$\begin{pmatrix}
	    -  & + \\ 
	    -  & +
	    \end{pmatrix}$ \\[5mm]
		
       $\chi_L$=$\begin{pmatrix}
	    \chi_L^{+} \\ 
	      \chi_L^{0}
	    \end{pmatrix}$  &1 &2 &1 &$\frac{1}{2}$ &0&0&$\begin{pmatrix}
	    + \\ 
	     +
	    \end{pmatrix}$  \\[5mm]
	    
       $\chi_R$=$\begin{pmatrix}
	    \chi_R^{+} \\ 
	    \chi_R^{0}
	    \end{pmatrix}$  &1 &1 &2 &$\frac{1}{2}$ &$\frac{1}{2}$&$\begin{pmatrix}
	    1 \\ 
	    0
	    \end{pmatrix}$ &$\begin{pmatrix}
	    - \\ 
	     +
	    \end{pmatrix}$  \\
    \hline\hline
   \end{tabular}
   \caption{The particle content of the ALRM with their respective gauge structure. The generalised lepton number is defined as $L=S+T_{3R'}$ with $S$ being $U(1)_S$ charge, and the $R$-parity as $R=(-1)^{3B+L+2s}$, where $B$ is the baryon number and $s$ as the spin.}
   \label{tab:my_label}
\end{table}
The scalar potential of the ALRM, which determines the Higgs spectrum and its mixings, is given by 
\begin{equation}\label{alrpot}
\begin{split}
 V_{\rm scalar} & = - \mu_1^2 ~{\rm Tr} \left[\Phi^{\dagger}\Phi \right] - \mu_2^2 \left(\chi_L^{\dagger}\chi_L +\chi_R^{\dagger}\chi_R \right) +   \lambda_1 \left( {\rm Tr} \left[\Phi^{\dagger}\Phi \right] \right)^2 + \lambda_2 {\rm Tr} \left[\Phi^{\dagger}\tilde{\Phi} \right]           {\rm Tr} \left[\tilde{\Phi}^{\dagger}
     \Phi \right] \\
     & +  \lambda_3 \left[ \left(\chi_L^{\dagger}\chi_L \right)^2 + \left(\chi_R^{\dagger}\chi_R \right)^2 \right] 
      +2\lambda_4 \left(\chi_L^{\dagger}\chi_L \right) \left(\chi_R^{\dagger}\chi_R \right) +2\alpha_1 {\rm Tr} \left[\Phi^{\dagger}\Phi \right] \left(\chi_L^{\dagger}\chi_L+\chi_R^{\dagger}\chi_R \right)\\
      &+ 2\alpha_2 \left[\chi_L^{\dagger}\Phi\Phi^{\dagger}\chi_L+\chi_R^{\dagger}\Phi^{\dagger}\Phi\chi_R \right] +2\alpha_3 \left[\chi_L^{\dagger}\tilde{\Phi} {\tilde{\Phi}}^{\dagger} \chi_L +\chi_R^{\dagger} {\tilde{\Phi}}^{\dagger} \tilde{\Phi}\chi_R \right]   +\mu_3 \left[\chi_L^{\dagger}\Phi\chi_R+\chi_R^{\dagger} \Phi^{\dagger} \chi_L \right],
 \end{split}
 \end{equation}
 where $\tilde{\Phi}=\sigma_2 \Phi^* \sigma_2$ and $\chi_{L,R}=i\sigma_2 \chi_{L,R}^*$. Here, we assume a left-right symmetry, and further consider all couplings to be real for simplicity. In the following, we shall  introduce the notation $\alpha_{12}=\alpha_1+\alpha_2$ and $\alpha_{13}=\alpha_1+\alpha_3$, acknowledging the fact that the operator in $\alpha_1$ term  can be expressed as sum of the operators in $\alpha_2$ and $\alpha_3$ terms.

The spontaneous breaking of the ALRM gauge symmetry down to electromagnetism proceeds in two steps, driven by a vacuum configuration in which all neutral scalar components, with the exception of $\phi_1^0$, acquire nonzero vacuum expectation values (VEVs):
\begin{equation}    
\langle\phi\rangle=\frac{1}{\sqrt{2}}
\begin{pmatrix}
    0 & 0 \\
    0 & v_u
\end{pmatrix}, \langle \chi_{L}\rangle=\frac{1}{\sqrt{2}}
\begin{pmatrix}
    0  \\
     v_{L}
\end{pmatrix}, ~~{\rm and}~~ \langle \chi_{R}\rangle=\frac{1}{\sqrt{2}}
\begin{pmatrix}
    0  \\
     v_R
\end{pmatrix}.
\end{equation}
The first step,  spontaneous breakdown of $SU(2)_{R'} \otimes U(1)_{B-L} \otimes U(1)_S$ to the SM gauge group proceeds in such a way that the generalized combination $L=S + T_{3R'}$, where $S$ is the $U(1)_S$ charge, remains unbroken. This is achieved through the neutral component of an $SU(2)_{R'} $ doublet $\chi_R$ charged under $U(1)_S$ acquiring a non-zero VEV $v_R$. The second step, the breaking of the electroweak symmetry down to electromagnetism, is driven by the VEVs $v_u$ and $v_L$ of the bi-doublet $\Phi$ and the $SU(2)_L$ doublet $\chi_L$, respectively. The vanishing of $\langle\phi_1^0\rangle$ is protected by the conservation of the generalized lepton number $L$, whereas $\phi_2^0$ and $\chi_L^0$ having $L=0$ are allowed to acquire non-zero VEVs.

Consequently, $\phi_1^0$ is prevented from mixing with any other scalar fields, resulting in two degenerate physical states, $H_1^0$ and $A_1$ corresponding the scalar and pseudo-scalar degrees of freedom of $\phi_1^0$ with mass:

\begin{equation} 
m^2_{H^0_{1}}=m^2_{A_{1}}=2v_u^2\lambda_{2}-(\alpha_{12}-\alpha_{13})(v_L^2+v_R^2)-\frac{v_L v_R \mu_3}{\sqrt{2}v_u}. \label{eq:mH10A1}
\end{equation}
The pseudo-scalar degrees of freedom of $\phi_2^0$, $\chi_L^0$, and $\chi_R^0$ mix together to become the two would be the Goldstone bosons absorbed by the neutral gauge bosons $Z_L$ and $Z_R$, along with the orthogonal combination manifesting as a physical pseudo-scalar particle $A_2$ with mass 
\begin{align}
   &   m_{A_2}^2=  - \frac{\mu_3}{\sqrt{2}} \frac{v_u^2 v_R^2+v_u^2 v_L^2+v_L^2 v_R^2}{v_Lv_R v_u}.          
 \end{align}
This immediately requires $\mu_3<0$.  Furthermore, from Eq.~\ref{eq:mH10A1}, the splitting $\alpha_{12} - \alpha_{13}$ is required to be at most $\mathcal{O}(10^{-2})$, as we shall explicitly demonstrate in the numerical analysis of the subsequent sections. Similarly, the scalar degrees of freedom of $\phi_2^0$, $\chi_L^0$, and $\chi_R^0$ mix together to produce three physical particles, denoted as $h,~H_2^0$, and $~H_3^0$. The lightest of these, $h$, emerges as the SM Higgs candidate, with its properties arranged to match those of the observed SM Higgs properties. The remaining two physical states, $H_2^0$ and $H_3^0$, are considerably heavier with masses being proportional to $v_R$.
For a detailed discussion of the scalar sector, we refer the reader to Ref.~\cite{Ashry:2013loa, Frank:2019nid}.
In the charged scalar sector, the physical massive charged Higgs bosons $H_1^\pm$ and $H_2^\pm$, together with the two massless Goldstone bosons $G_1^\pm$ and $G_2^\pm$ which are absorbed by the $W_L$ and $W_R$ gauge bosons, respectively, are given by
\begin{equation}
\begin{pmatrix}
\phi_1^\pm \\ \chi_L^\pm
\end{pmatrix}
=
\begin{pmatrix}
\cos\beta & \sin\beta \\
-\sin\beta & \cos\beta
\end{pmatrix}
\begin{pmatrix}
H_1^\pm \\ G_1^\pm
\end{pmatrix},
\quad\quad
\begin{pmatrix}
\phi_2^\pm \\ \chi_R^\pm
\end{pmatrix}=
\begin{pmatrix}
\cos\zeta & \sin\zeta \\
-\sin\zeta & \cos\zeta
\end{pmatrix}
\begin{pmatrix}
H_2^\pm \\ G_2^\pm
\end{pmatrix},
\end{equation}
where $\tan\beta=\frac{v_u}{v_L}$ and
$\tan\zeta=\frac{v_u}{v_R}$.
The masses of the charged Higgs bosons
$H_1^\pm$ ($R$-parity even) and
$H_2^\pm$ ($R$-parity odd) are given by
\begin{equation}
m^2_{H_1^\pm} =- v^2\left[(\alpha_{12}-\alpha_{13})+\frac{\mu_3 v_R}{\sqrt{2}\,v_u v_L}\right], \label{eq:mH1p}
\end{equation}
and
\begin{equation}
m^2_{H_2^\pm} =- v'^2\left[(\alpha_{12}-\alpha_{13})+\frac{\mu_3 v_L}{\sqrt{2}\,v_u v_R}\right], \label{eq:mH2p}
\end{equation}
  respectively, where $v^2 = v_u^2+v_L^2$ and $v'^2=v_R^2+v_u^2$. This again required $(\alpha_{12}-\alpha_{13})<{\cal O}(10^{-2})$, leaving the second term proportional to $\mu_3$ as the dominant one.
Since $v'\gg v$, and in most scenarios we consider $v_u\gg v_L$, the charged Higgs, $H_1^\pm$ is typically heavy, while $H_2^\pm$ lies in the sub-TeV to TeV range over a significant portion of the model parameter space. For appropriate parameter choices, the light charged Higgs boson can be long-lived, with its decay potentially yielding a disappearing track signature at the LHC~\cite{Deka:2026gfo}, which is a distinctive collider signal of the model.

In the gauge boson sector, the breaking of left-right symmetry generates masses 
for the gauge bosons and induces mixing among them. Due to the conservation of 
the generalized lepton number, which prevents $\langle\phi_1^0\rangle$ from 
being nonzero, the charged $W_L$ and $W_R$ bosons do not mix. The masses of 
these gauge bosons are given by
\begin{equation}
    M^2_{W_L} = \frac{g_L^2}{4}\left(v_u^2 + v_L^2\right) 
    ,\qquad{\rm and}\quad 
    M^2_{W_R} = \frac{g_R^2}{4}\left(v_u^2 + v_R^2\right).  
    \label{eq:MWRmass}
\end{equation}
The charge-neutral gauge bosons, however, do undergo mixing, and the 
diagonalization of the mixing matrix leads to the physical mass eigenstates 
$(A_\mu,\, Z_\mu,\, Z'_\mu)$. The $Z_L$-$Z_R$ mixing in $Z'$ is strongly constrained 
experimentally, and in the limit where this mixing vanishes, the mass eigenvalues 
are obtained analytically, with the photon remaining massless. The masses of $Z$ 
and $Z'$ are given by
\begin{equation}
    M^2_Z = \frac{g^2_L (v_u^2+v_L^2)}{4\cos^2\theta_W},
    \qquad{\rm and} \qquad
    M^2_{Z'} = \tfrac{1}{4}g_{BL}^2 \sin^2\phi_W\, v_L^2 
    + \frac{g_R^2\left(\cos^4\phi_W\, v_u^2 + v_R^2\right)}{4\cos^2\phi_W},
    \label{eq:MZmass}
\end{equation}
respectively, where $g_L$, $g_R$, and $g_{BL}$ denote the gauge couplings corresponding to 
$SU(2)_L$, $SU(2)_{R'}$, and $U(1)_{B-L}$, respectively, $\theta_W$ is the 
Weinberg mixing angle, and $\phi_W$ is the corresponding mixing angle in the 
right-handed gauge sector. In the hierarchy $v_R \gg v_u \gg v_L$, the term 
involving $g_R$ dominates the $Z'$ mass, and one can reasonably approximate 
$M_{Z'} \approx M_{W_R}/\cos\phi_W$.
Since $W_R$ does not decay exclusively to the SM particles, it evades direct search constraints at colliders. The $Z'$, on the other hand, is subject to direct search bounds from the collider experiments, which translate into indirect constraints on 
$W_R$ through the above relation. For a detailed discussion of the LHC constraints 
on the gauge bosons of the ALRM, we refer the reader to Ref.~\cite{Frank:2024imi}.

Finally, turning to the Yukawa interactions in the ALRM framework, we have 
\begin{eqnarray}\label{Yukawa}
- \mathcal{L}_Y &=&
      Y_{Q1} \, \overline{Q}_{L} \chi_L d_{R}
    + Y_{Q2} \, \overline{Q}_{L} \tilde{\Phi} Q_{R}
    + Y_{Q3} \, \overline{Q}_{R} \chi_R d'_{L}
\nonumber \\
    &&+ Y_{L1} \, \overline{L}_{L} \Phi L_{R}
    + Y_{L2} \, \overline{L}_{L} \tilde{\chi}_L \nu_{R}
    + Y_{L3} \, \overline{L}_{R} \tilde{\chi}_R n_{L}
    + \mathrm{h.c.},
\end{eqnarray}
where the flavor indices are suppressed, and the dual fields are defined as $\tilde{\Phi} = \sigma_2 \Phi^\ast \sigma_2$ and $\tilde{\chi}_{L,R} = i\sigma_2 \chi_{L,R}^\ast$ with $\sigma_2$ being the second Pauli matrix. 
With the VEVs of scalars being as aforementioned, the fermion mass matrices are given by
\begin{align}
    &M_{u}=\frac{v_u}{\sqrt{2}} Y_{Q2},~~~M_{d}=\frac{v_L}{\sqrt{2}} Y_{Q1},~~~ M_{\ell}=\frac{v_u}{\sqrt{2}} Y_{L1}, \nonumber
    \\&M_{\nu}=\frac{v_L}{\sqrt{2}} Y_{L2},~~~M_{d'}=\frac{v_R}{\sqrt{2}} Y_{Q3},~~~M_n=\frac{v_R}{\sqrt{2}} Y_{L3}.
\end{align}

In the present paper, our focus is the scalar potential and the theoretical constraints arising from the considerations of the vacuum stability and the perturbative unitarity. We devote the remainder of the article to this discussion.

\section{Theoretical Constraints}
\label{sec:theoretical}
In this section, we discuss the theoretical constraints on the parameter space of the ALRM framework, starting with the tree-level vacuum stability constraints, followed by the constraints imposed by the perturbative unitarity.

\subsection{Vacuum stability constraints}\label{vacuum_stability}

\begin{figure}[h]
\centering
\includegraphics[width=0.465\textwidth]{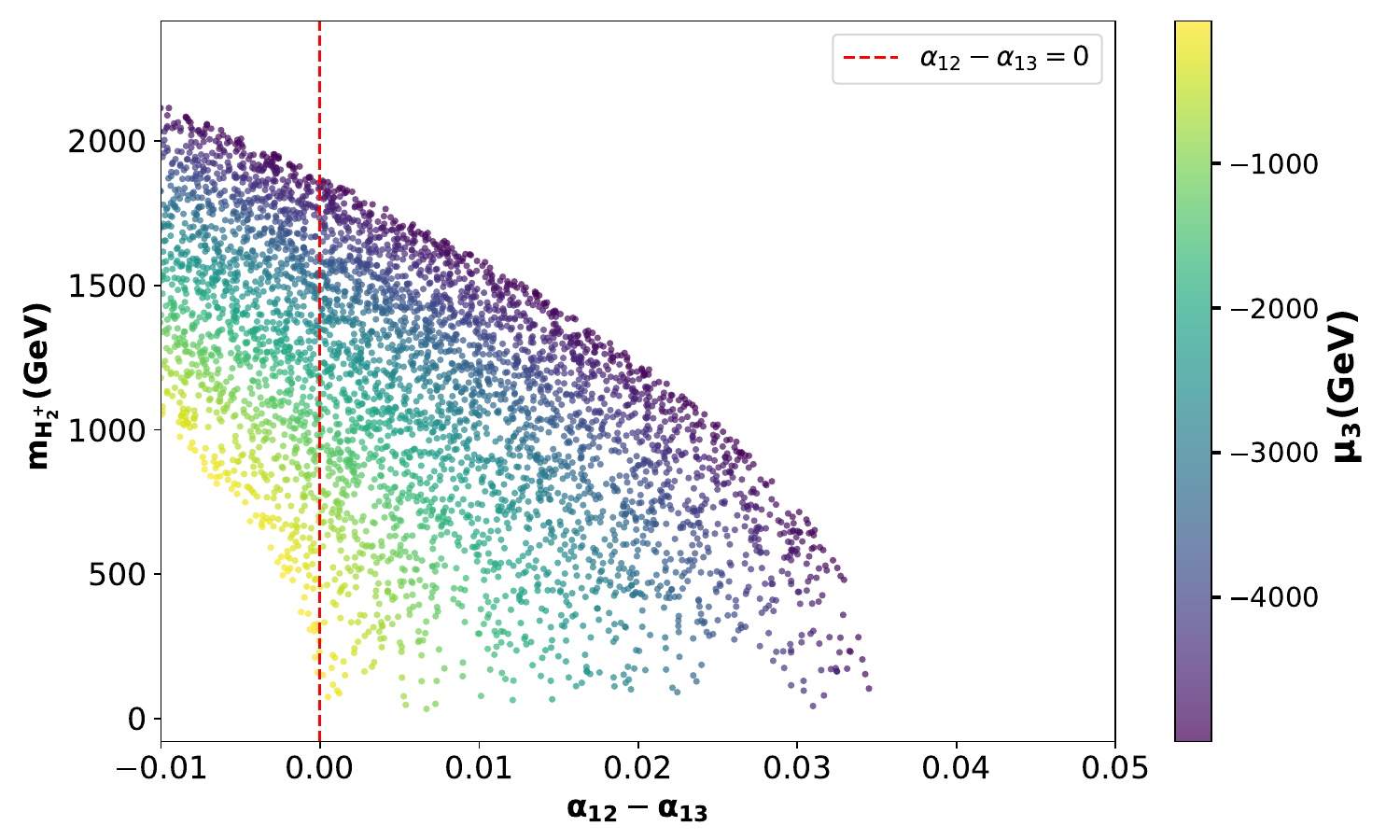}
\includegraphics[width=0.465\textwidth]{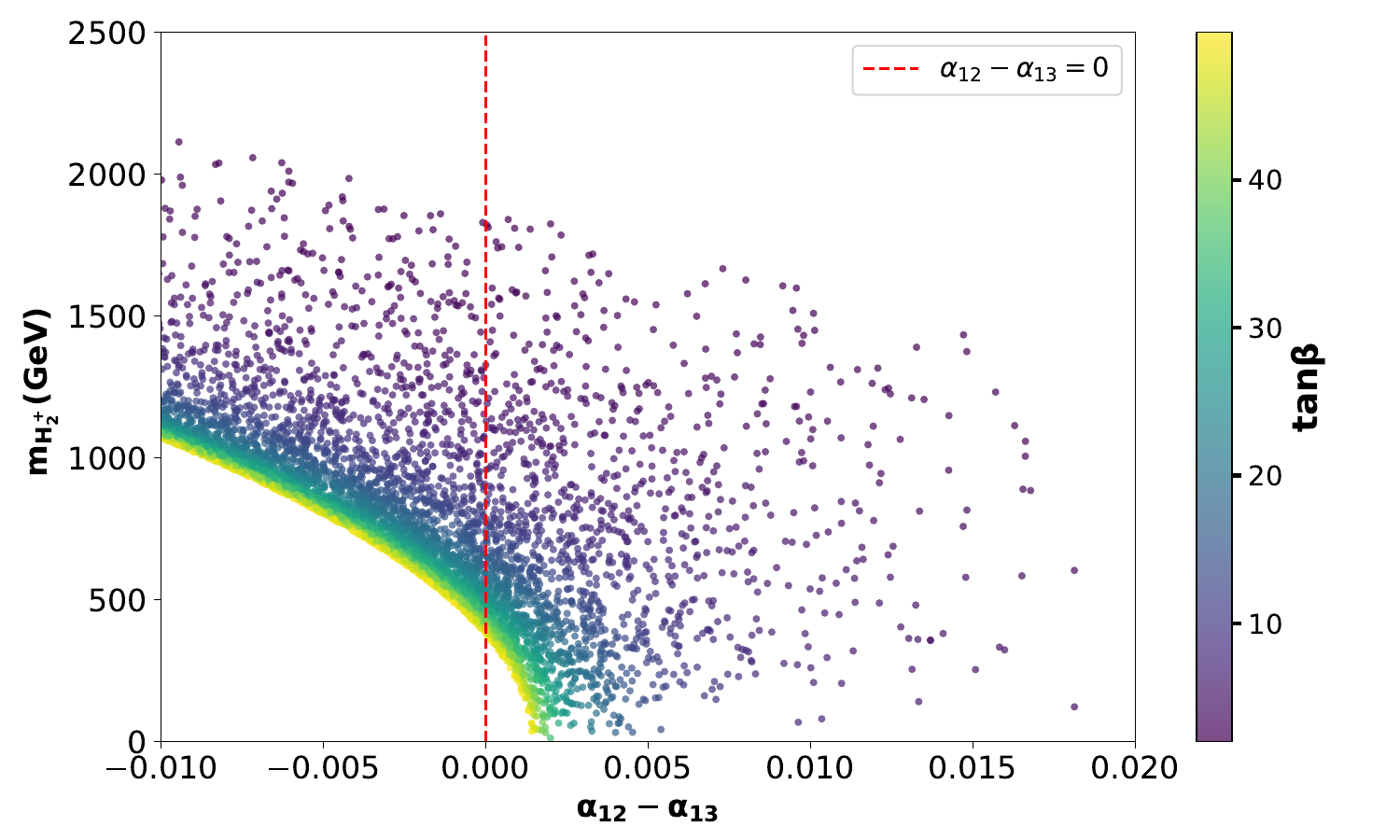}
\caption{\label{alp23vsmH}Dependence of the charged Higgs mass $m_{H^\pm_2}$ on $(\alpha_{12}-\alpha_{13})$. ({\it Left}): The impact of $\mu_3$ for fixed $\tan\beta=10$ and $v_R=10$ TeV; ({\it Right}): The sensitivity of $\tan\beta$ for fixed $\mu_3 = -1000$ GeV and $v_R=10$ TeV. The negative values of $(\alpha_{12} - \alpha_{13})$ are disallowed for the potential to be bounded from below.}
\end{figure}

The stability of the vacuum of the ALRM scalar potential has been investigated at tree level in Ref.~\cite{Frank:2021ekj} using copositivity criteria. For the scalar potential to be bounded from below, it must remain positive in all field directions as the fields tend to infinity. This immediately requires the quartic couplings involving only a single field to be positive definite. However, the couplings corresponding to mixed quartic terms involving more than one field may be negative, provided the negative contribution is compensated by the single-field quartic couplings in all field directions. To illustrate this, note that the operator multiplying $\lambda_2$ can be expressed in terms of components as \( {\rm Tr} \left[\Phi^{\dagger}\tilde{\Phi} \right]           {\rm Tr} \left[\tilde{\Phi}^{\dagger}
     \Phi \right]=|\phi_1^{0}\phi_2^{0}-\phi_1^+\phi_2^-|^2\), 
which involves products of two distinct fields, and whose negative contribution can be compensated by the single-field quartic terms such as $(\phi_1^0)^4$ and $(\phi_2^0)^4$,
thus permitting $\lambda_2<0$.
This leads to the copositivity conditions, which are necessary and sufficient conditions for the scalar potential to be bounded from below. A detailed and systematic derivation of these conditions for the ALRM scalar potential is presented in Ref.~\cite{Frank:2021ekj}. The constraints arising from the copositivity conditions, combined with the minimization conditions for the desired global vacuum, impose the following restrictions on the quartic couplings:
\vskip 3mm
\hrule
\vspace{-2mm}
\begin{minipage}{.3\textwidth}
\begin{equation} \label{BFB_conditions}
\begin{aligned}
    &\lambda_1\ge 0,~~\lambda_4 \geq \lambda_3 \geq0,~~~~~~~~~~  \\
&\lambda_1+2\lambda_2\geq0,~~\lambda_2\leq 0,\\
&\alpha_{12}- \alpha_{13}\geq0, \nonumber
\end{aligned}
\end{equation}
\end{minipage}\hfill
\begin{minipage}{.6\textwidth}
\begin{equation} \label{BFB_conditions}
\begin{aligned}
    &\alpha_{12} + \sqrt{\lambda_1 \left( \frac{\lambda_3 + \lambda_4}{2} \right)} \geq 0, ~~~
    \alpha_{12} + \sqrt{\lambda_1 \lambda_3} \geq 0, \\
    &\alpha_{13}+ \sqrt{\lambda_1 \left( \frac{\lambda_3 + \lambda_4}{2} \right)} \geq 0, ~~
    \alpha_{13}  + \sqrt{\lambda_1 \lambda_3} \geq 0.
\end{aligned}
\end{equation}
\end{minipage}
\vskip 2mm
\hrule
\vskip 5mm
A remark regarding $(\alpha_{12}-\alpha_{13})$ is in order here. The copositivity 
condition requires $(\alpha_{12}-\alpha_{13})\geq 0$, while the positivity of 
$m^2_{H_2^\pm}$ naively prefers $(\alpha_{12}-\alpha_{13})\leq 0$ 
(Eq.~\ref{eq:mH2p}). On this basis, Ref.~\cite{Frank:2021ekj} concludes that 
$\alpha_{12}-\alpha_{13}=0$. We investigate this further and find that a small 
positive value of $(\alpha_{12}-\alpha_{13})$ is in fact consistent with 
$m^2_{H_2^\pm} > 0$, its allowed range depending on the other parameters of the 
model, in particular on $\mu_3$ and $\tan\beta$. To demonstrate this, in 
Fig.~\ref{alp23vsmH}, we plot $m_{H_2^\pm}$ as a function of $(\alpha_{12}-\alpha_{13})$ 
for fixed $\tan\beta=10$ and $v_R=10~\text{TeV}$, scanning $\mu_3$ randomly over 
the range $[-5000,~-10]~\text{GeV}$ (left panel), and for fixed $\mu_3=-1000~\text{GeV}$ 
and $v_R=10~\text{TeV}$, varying $\tan\beta$ between 2 and 50 (right panel). 
Negative values of $(\alpha_{12}-\alpha_{13})$ are excluded by the 
boundedness-from-below condition. It is evident from the figure that smaller 
values of $\tan\beta$ and larger $|\mu_3|$ permit a larger positive splitting 
$(\alpha_{12}-\alpha_{13})$.

\subsection{Perturbative unitarity constraints}\label{sec_unitarity}
At tree level, another theoretical constraint on the ALRM parameter space emanates from the perturbative unitarity of the $2\to 2$ scalar scattering amplitudes in the high-energy limit. It demands that the zeroth partial wave amplitude $a_0$ must satisfy $|{\rm Re}(a_0)| \le \frac{1}{2}$ \cite{Krauss:2017xpj}.
In general, the invariant amplitude of the $2\to2$ scattering process from the partial wave decomposition could be given by
\begin{equation}
    {\cal{M}}(s,\theta)= 16 \pi~ \sum^{\infty}_{J=0} a_J(s)(2J+1) ~P_J(cos ~\theta), 
\end{equation}
where $P_J$ denotes the $J^{\rm th}$ degree Legendre polynomial with $J$ denoting the total orbital angular momentum of the final state. In 
addition, $\sqrt{s}$ denotes the center-of-mass energy and $\theta$ is the scattering angle. For the $2\to 2$ scalar scattering amplitudes being real at tree level, it is optimal to use the strong condition $|{\rm Re}(a_0)| \le \frac{1}{2}$ which leads to $| \mathcal{M}| \le 8\pi$. Finally, we obtain the perturbative unitarity constraints for the model framework by demanding all the eigenvalues of $\mathcal{M}$ be  $\le 8\pi$.

Given the extended scalar sector of the ALRM framework, characterized by nine independent parameters in the scalar potential, the constraints arising from the perturbative unitarity are highly non-trivial. In total, there are 136 independent $2 \to 2$ scattering processes involving scalar fields in the model. These processes can be systematically classified according to the total electric charge of the initial state: the charge-neutral, the singly charged, and the doubly charged sectors. We construct the scattering matrix $\mathcal{M}$ in the basis of all possible two-particle states and arrange it in a block-diagonal form, where each block corresponds to a definite charge sector. The full matrix decomposes into the neutral charged block of dimension 52, the singly charged blocks of dimension 32 each (for the positive and the negative charges, related by Hermitian conjugation), and the doubly charged blocks of dimension 10 each (again related by Hermitian conjugation). Altogether, these blocks account for the full 136-dimensional scattering matrix. 
Rotating to the diagonal basis, one obtains 136 eigenvalues corresponding to the coupled-channel system. Owing to the symmetry structure of the scalar potential, there is significant degeneracy among the eigenvalues of the scalar scattering matrix, and only 14 independent constraints from the perturbative unitarity were obtained \footnote{The Mathematica files containing the scattering matrices and the corresponding eigenvalues are available at \url{https://github.com/hdeka1/Unitarity-constraints-file}}:
\begin{align}
    e_{1,2} &:~~~~ |\alpha_{12,~13}| \le 4\pi, \label{eq:eigen1n2}\\
    e_{3,4,5} &:~~~~ |\lambda_{1,3,4}| \le 4\pi, \label{eq:eigen3n4n5} \\
    e_6 &:~~~~ |\alpha_{12} - 2\alpha_{13}| \le  4\pi,\label{eq:eigen6} \\
    e_7 &:~~~~ |2\alpha_{12} - \alpha_{13}| \le 4\pi, \label{eq:eigen7}\\
    e_8 &:~~~~ |\lambda_1 + 2\lambda_2| \le 4\pi, \label{eq:eigen8}\\
    e_9 &:~~~~ |\lambda_1 + 4\lambda_2| \le 4\pi, \label{eq:eigen9}\\
    e_{10} &:~~~~ |2\lambda_4 - 3\lambda_3| \le  4\pi,\label{eq:eigen10} \\
    e_{11,12} &:~~~~ \left|\lambda_1 - 2\lambda_2 + \lambda_3 \pm 
    \sqrt{(\lambda_1 - 2\lambda_2 - \lambda_3)^2 + 
    8(\alpha_{12}-\alpha_{13})^2}\,\right| \le 8\pi, \label{eq:eigen11n12}\\
    e_{13,14} &:~~~~ \left|5\lambda_1 + 2\lambda_2 + 3\lambda_3 + 2\lambda_4 \pm 
    \sqrt{(5\lambda_1 + 2\lambda_2 - 3\lambda_3 - 2\lambda_4)^2 + 
    16(\alpha_{12}+\alpha_{13})^2}\,\right| \le  8\pi. \label{eq:eigen13n14}
\end{align}

The independent unitarity constraints on the quartic couplings are labeled as $e_i$ ($i = 1, \ldots, 14$) for ease of reference. For the paired eigenvalues, $e_{11}$ and $e_{13}$ correspond to the positive branch of the square root, while $e_{12}$ and $e_{14}$ correspond to the negative branch. These results have been independently verified using \texttt{SARAH}~\cite{Staub:2015kfa}, a general-purpose framework for studying BSM models, and \texttt{anyPUB}~\cite{Benincasa:2025tfq}, a dedicated package for deriving perturbative unitarity constraints.

\subsection{Numerical Study}
\label{sec:theroeticalscan}
We perform a random scan over the parameter space to identify regions consistent with the vacuum stability, the perturbative unitarity, and the requirement that all physical scalar mass-squared eigenvalues are positive. Of the nine couplings in the scalar potential, four parameters ($\mu_1,~\mu_2,~\lambda_1,~\lambda_4$) are traded for the vacuum expectation values $v_R$, $v_u = v\sin\beta$, $v_L = v\cos\beta$ (see Appendix~\ref{tadpole}), and the SM-like Higgs mass $m_h$. This leaves six independent parameters:
\begin{equation}
\tan\beta,\quad \lambda_2,\quad \lambda_3,\quad \alpha_{12},\quad 
\alpha_{13},\quad {\rm and}\quad \mu_3.
\end{equation}

In our scan, we fix $v_R = 10~\text{TeV}$ and vary the remaining parameters over the ranges given as:
\begin{equation}\label{uni_scan}
\begin{aligned}
\alpha_{12},~\alpha_{13} &\in [0,~4\pi], \\
\lambda_2 &\in [-4\pi,~0], \\
\lambda_3 &\in [0,~4\pi], \\
\tan\beta &\in [2,~50], \\
\mu_3 &\in [-10^4,~-10]~\text{GeV}.
\end{aligned}
\end{equation}

The lower bound $\tan\beta = 2$ is chosen to reproduce the top quark mass with a Yukawa coupling $Y_{Q2} \sim 1$, using $m_t = \frac{1}{\sqrt{2}} Y_{Q2}\, v\tan\beta \simeq 172.5~\text{GeV}$. The upper bound $\tan\beta = 50$ corresponds to $v_L \sim 5~\text{GeV}$, ensuring that the bottom quark mass can be generated with a Yukawa coupling of order unity. The range of $\mu_3$ is selected such that the charged Higgs mass $m_{H_2^\pm}$ lies within $100~\text{GeV} \text{--} 2000~\text{GeV}$. The remaining couplings are scanned over the full range allowed by the perturbativity. The parameters $\tan\beta$ and $\mu_3$ play a special role through their implicit appearance in $\lambda_1$ and $\lambda_4$, which are determined via the scalar mass spectrum. Since this dependence involves the neutral scalar mass eigenstates in a non-trivial manner, we refer the reader to Refs.~\cite{Ashry:2013loa, Frank:2019nid, Frank:2021ekj} for more details.

Throughout this work, we fix $v_R = 10~\text{TeV}$ unless stated otherwise. This is justified because all parameters entering the theoretical constraints, except the derived quantities $\lambda_1$ and $\lambda_4$, are independent of $v_R$. Moreover, the dependence of $\lambda_1$ and $\lambda_4$ on $v_R$ always appears in combination with $\mu_3$, and its effect is largely absorbed by scanning over $\mu_3$. We have explicitly verified that varying $v_R$ does not significantly affect the resulting constraints.
\begin{figure}[h]
\includegraphics[width=1\textwidth]{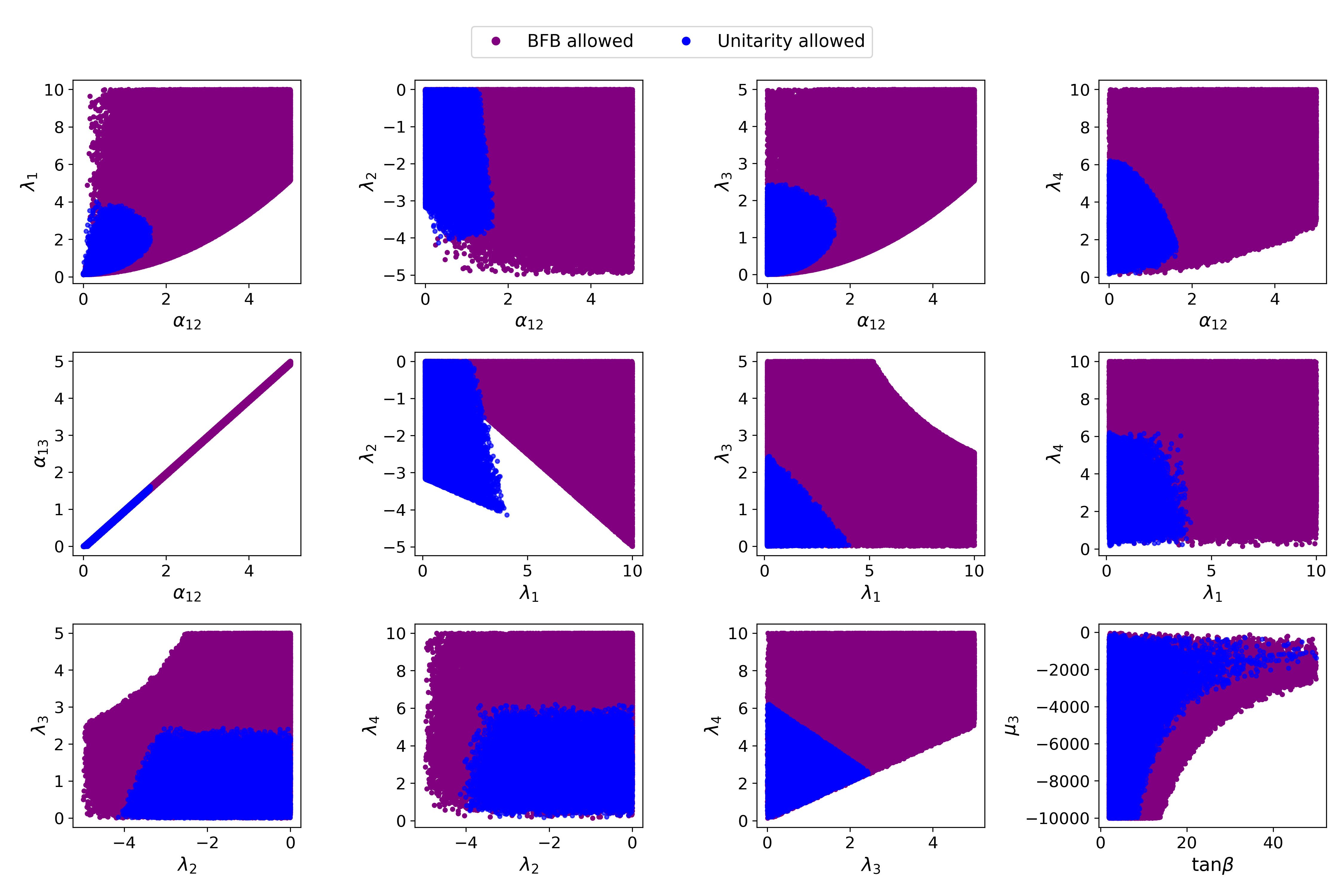}
\caption{\label{BFB_Unitarity}Correlation plots showing the allowed region by the perturbative unitarity (blue) and the BFB conditions (purple). The scan range as specified in Eq. \ref{uni_scan}, and we kept $v_R$=10 TeV. In both cases, 
the positivity of all scalar mass-squared eigenvalues is maintained. This leads to absolute upper limits on the couplings: $(\alpha_{12},~\alpha_{13})\lesssim 1.4,~~\lambda_1\lesssim 3.0,~~|\lambda_{2}|\lesssim 1.5,~~\lambda_{3}\lesssim 2.5, {\rm ~and }~\lambda_4\lesssim 2\pi$.}
\end{figure}
The results of the scan over one million ($10^6$) points are presented in 
Fig.~\ref{BFB_Unitarity}, showing the correlations among the parameters. Points 
satisfying the vacuum stability conditions of Eq.~\ref{BFB_conditions} are 
shown in purple, while those satisfying the perturbative unitarity bounds of 
Eqs.~\ref{eq:eigen1n2}--\ref{eq:eigen13n14} are shown in blue. In both cases, we have made sure that the mass-square relations are positive definite. The two sets of 
constraints are complementary: each excludes regions of parameter space not ruled 
out by the other, with a partial overlap in the allowed regions. This underscores 
the importance of imposing all theoretical constraints simultaneously, as 
considering them in isolation would significantly overestimate the viable parameter 
space of the model. The strong correlation between $\alpha_{12}$ and $\alpha_{13}$ 
is a direct consequence of the vacuum stability and mass-squared positivity 
conditions, which together restrict the splitting $(\alpha_{12}-\alpha_{13})$ to 
small positive values, as discussed above.
A strong correlation between $\alpha_{12}$ and $\alpha_{13}$ is observed, which can be understood from the vacuum stability and the positivity conditions that permit only a small positive splitting $(\alpha_{12}-\alpha_{13})$, as discussed above.
The scan results indicate approximate upper bounds on the quartic couplings,
\begin{equation}
(\alpha_{12},~\alpha_{13})\lesssim 1.4, \qquad 
\lambda_1\lesssim 3.0, \qquad 
|\lambda_{2}|\lesssim 1.5, \qquad 
\lambda_{3}\lesssim 2.5, \qquad {\rm and }\qquad
\lambda_4\lesssim 2\pi,
\label{eq:ThConstraints}
\end{equation}
which arise from the combined effect of the perturbative unitarity, the vacuum stability, and the positivity of scalar mass-squared eigenvalues.

Within this allowed region, the parameters are further restricted by correlations. For instance, $\lambda_4 \le 2\pi$ when $\lambda_3 = 0$, while Eq.~\ref{eq:eigen13n14} implies the relation
\(
3\lambda_3 + 2\lambda_4 \lesssim 4\pi,
\)
leading to an upper bound
\begin{equation}
\lambda_4^{\mathrm{max}} \simeq 2\pi - \tfrac{3}{2}\lambda_3,
\end{equation}
which is clearly reflected in the $\lambda_3$--$\lambda_4$ correlation plot in Fig.~\ref{BFB_Unitarity}. 
In addition, the vacuum stability requires $\lambda_4 \ge \lambda_3$ (Eq.~\ref{BFB_conditions}), which further constrains $\lambda_3 \lesssim 2.5$ for the chosen ranges of $\mu_3$, $\tan\beta$, and $v_R$, with a corresponding value of $\lambda_4\sim 2.5$. A pronounced correlation between $\tan\beta$ and $\mu_3$ is also evident from Fig.~\ref{BFB_Unitarity}. This inverse correlation arises from the relation $\lambda_4 - \lambda_3 = -\tfrac{\mu_3 \tan\beta}{\sqrt{2} v_R}$, together with the perturbative bounds on $\lambda_3$ and $\lambda_4$.

Overall, the perturbative unitarity constraints, derived here for the first time in this framework, significantly reduce the allowed parameter space beyond what is implied by the vacuum stability considerations alone.



\section{Renormalization Group Evolution}\label{sec:RGE}

The vacuum stability and the perturbative unitarity conditions derived in the previous section hold at tree level and at a fixed energy scale. To assess the theoretical consistency of the model across a wider range of energies, these conditions must be re-examined using the Renormalization Group Evolution (RGE) of the couplings. In the limit of large field values, the scalar potential is well approximated by its tree-level form with couplings replaced by their running counterparts, $C_i \to C_i(Q)$. The full set of running parameters comprises:
\begin{equation}
C_i \in \left\{
\begin{array}{ll}
\text{gauge couplings:} & g_3,~g_L,~g_R,~g_{BL} \\[4pt]
\text{Yukawa couplings:} & Y_{Q1},~Y_{Q2},~Y_{Q3},~Y_{L1},~Y_{L2},~Y_{L3} \\[4pt]
\text{quadratic \& trilinear couplings:} & \mu_1,~\mu_2,~\mu_3 \\[4pt]
\text{quartic couplings:} & \lambda_1,~\lambda_2,~\lambda_3,~\lambda_4,~\alpha_{12},~\alpha_{13}.
\end{array}
\right.
\end{equation}
Not all of these are independent. As discussed in Sec.~\ref{sec:theroeticalscan}, the parameters $\mu_1$, $\mu_2$, and $\lambda_4$ are determined by the tadpole conditions (see Appendix~\ref{tadpole}), and $\lambda_1$ is fixed in terms of the remaining parameters together with the requirement that the lightest CP-even scalar reproduces the observed SM Higgs mass, $m_h = 125$~GeV. The independent input parameters for the RGE analysis are therefore the gauge couplings, specified at the electroweak scale as discussed in Sec.~\ref{subsec:gaugecouplings}; the Yukawa couplings, expressed in terms of the physical fermion masses and the vacuum expectation values; and the scalar potential parameters
\begin{equation}
\lambda_2, \quad \lambda_3, \quad \alpha_{12}, \quad \alpha_{13}, \quad \mu_3,
\end{equation}
all specified at the electroweak scale $v = 246$~GeV.

The RGEs of the model parameters are obtained by implementing the ALRM in \texttt{PyR@TE}~\cite{Lyonnet:2013dna, Lyonnet:2015jca}, and the complete set of $\beta$-functions is collected in Appendix~\ref{app:RGE}. The RG running is implemented in two stages, reflecting the two-step symmetry breaking pattern of the model. Below the scale $v_R$, the SM $\beta$-functions govern the running of the gauge and Yukawa couplings from $v$ up to $v_R$, while the ALRM-specific scalar couplings are held fixed, since they decouple from the SM running at this stage. At $v_R$, the gauge couplings are matched across the symmetry breaking threshold $\mathrm{SU}(2)_{R'} \otimes {U}(1)_{B-L} \to {U}(1)_Y$, and the full ALRM $\beta$-functions are then used to evolve all couplings to higher energies.

At each scale above $v_R$, we verify simultaneously that: (i)~the vacuum stability conditions of Eq.~\eqref{BFB_conditions} are satisfied; (ii)~all the perturbative unitarity bounds of Eqs.~\eqref{eq:eigen1n2}--\eqref{eq:eigen13n14} are respected; and (iii)~all the physical scalar mass-squared eigenvalues remain positive definite. The scale at which any one of these conditions is first violated defines a theoretical upper bound on the validity of the perturbative description. In parallel, we track the onset of any Landau pole, where a coupling diverges and the perturbative expansion breaks down entirely. The cutoff scale of the theory is then identified as the lesser of the scale at which a theoretical constraint is violated and the scale at which a Landau pole is encountered, whichever is reached first signals the boundary beyond which the model can no longer be considered perturbatively reliable without the introduction of new dynamics.

\subsection{Gauge Couplings}
\label{subsec:gaugecouplings}
We shall first look at the running of the gauge couplings. Starting at the electroweak scale, we consider the $g_L$ and $g_{Y}$ running up to $v_R$. Beyond this, we have the $g_Y$ given in terms of $g_R$ and $g_{BL}$:
\begin{equation} \label{gygRgBL}
    \frac{1}{g_Y^2}=\frac{1}{g_{BL}^2}+\frac{1}{g_R^2}.
\end{equation}
To fix the freedom in the choice of gauge couplings, we relate $g_R$ to $g_L$, and define their ratio as 
\begin{equation}
r_g \equiv \tfrac{g_R(v_R)}{g_L(v_R)}
\end{equation}
Neglecting the mixing between $Z_L-Z_R$, the ratio between the heavy gauge boson masses is given by
\begin{equation}\label{zpwp}
    \frac{M^2_{Z^{'}}}{M^2_{W_{R}}}\approx  r^2_g\Big(   r_g^2-\text{tan}^2\theta_W      \Big),
\end{equation}
where $\theta_W$ being the weak mixing angle. It results in providing an upper limit for the ratio $r_g$ as the ratio of the mass squares of gauge bosons must always remain a non-negative value. Henceforth, we get 
\begin{equation}
    r_g \geq \text{tan}~\theta_W \approx 0.55.
\end{equation}
The other coupling $g_{BL}$ is then determined at the scale $v_R$ through the matching condition of Eq.~\eqref{gygRgBL},
\begin{equation}
   \frac{1}{ g^2_{BL}(v_R)}
    = \frac{1}{g^2_Y(v_R)}-\frac{1}{r_g^2\,~ g^2_L(v_R)},
\end{equation}
which ensures consistency with the hypercharge relation at the symmetry breaking threshold. The ratio $r_g$ is a free parameter of the matching procedure, subject to the requirement that the resulting values of $g_R$ and $g_{BL}$ remain within the perturbative regime not only at $v_R$ but throughout their evolution up to the cutoff scale of the model.

Above $v_R$, the full ALRM gauge symmetry is restored and the couplings $g_{BL}$ and $g_R$ evolve independently according to their respective one-loop $\beta$-functions:
\begin{align}
    16\pi^2\,\beta(g_{BL}) &= \frac{13}{3}\,g_{BL}^3, \\[6pt]
{\rm  and }\nonumber\\    16\pi^2\,\beta(g_{R}) &= -\frac{17}{6}\,g_{R}^3.
\end{align}
Here, the opposite signs of the $\beta$-functions reflect qualitatively different high-energy 
behaviours of the two couplings: $g_R$ is asymptotically free and decreases monotonically with increasing energy scale, while $g_{BL}$ grows and may eventually approach its perturbative limit. This asymmetry, which is a direct consequence of the non-abelian versus the abelian nature of the respective gauge groups, plays an important role in determining the allowed range of $r_g$ and hence the viable parameter space of the model at high energies. We examine the evolution of all gauge couplings up to energies of order $10^{16}$~GeV, which we use as a representative high energy cut-off scale and refer to as the grand unified theory (GUT) scale throughout the article for convenience. We emphasize, however, that gauge coupling unification is not imposed as a  requirement of the model.

For numerical analysis, we consider the values of the couplings $g_L$ and $g_Y$ as \cite{PhysRevD.110.030001}:
\[
g_L(M_Z) =0.65096 ~\pm~0.00004 ,~~~{\rm and }~~~g_Y(M_Z) = 0.35726~\pm~0.00002.
\]
\begin{figure}[h]
    \centering
    \begin{subfigure}[b]{0.48\textwidth}
        \centering
        \includegraphics[width=\linewidth]{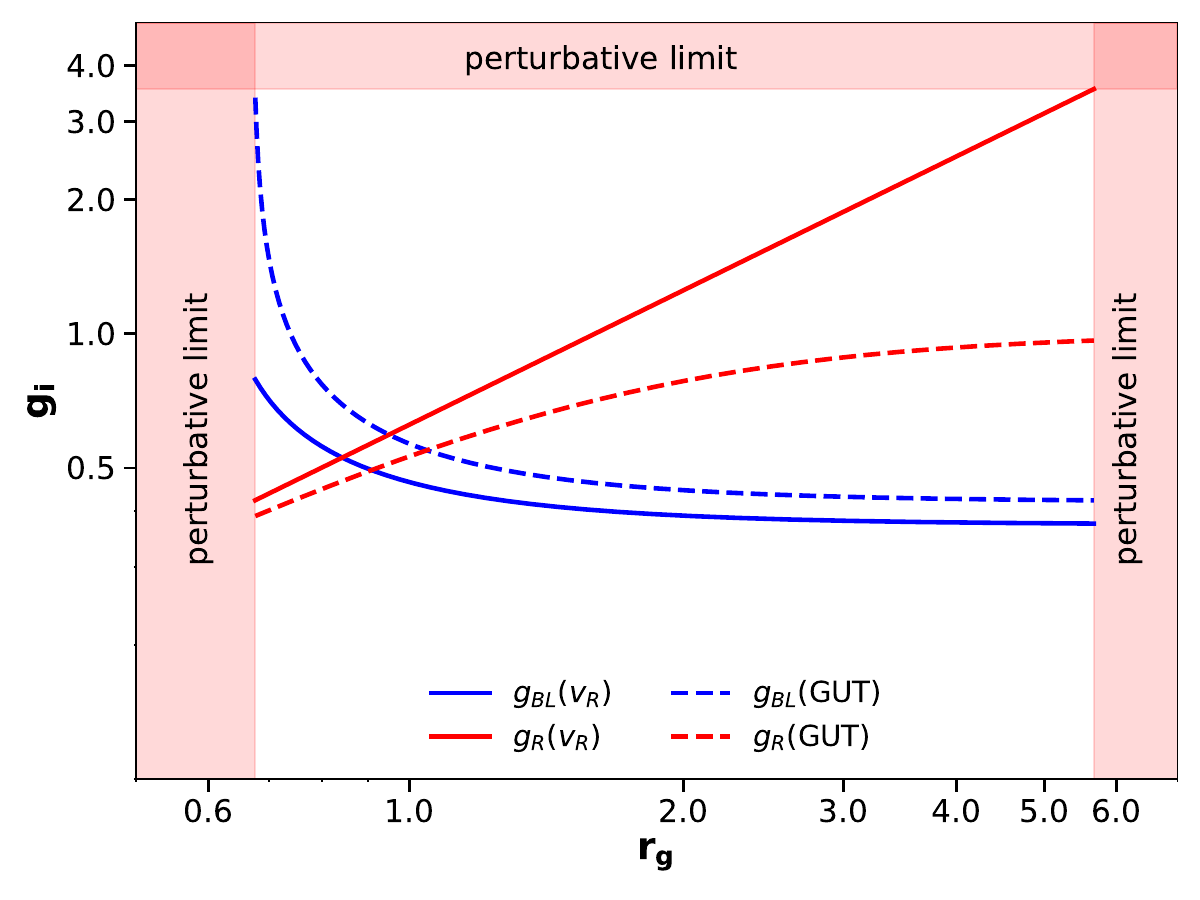}
        \caption{}
    \end{subfigure}\hspace{5mm}
    \begin{subfigure}[b]{0.48\textwidth}
        \centering
        \includegraphics[width=\linewidth]{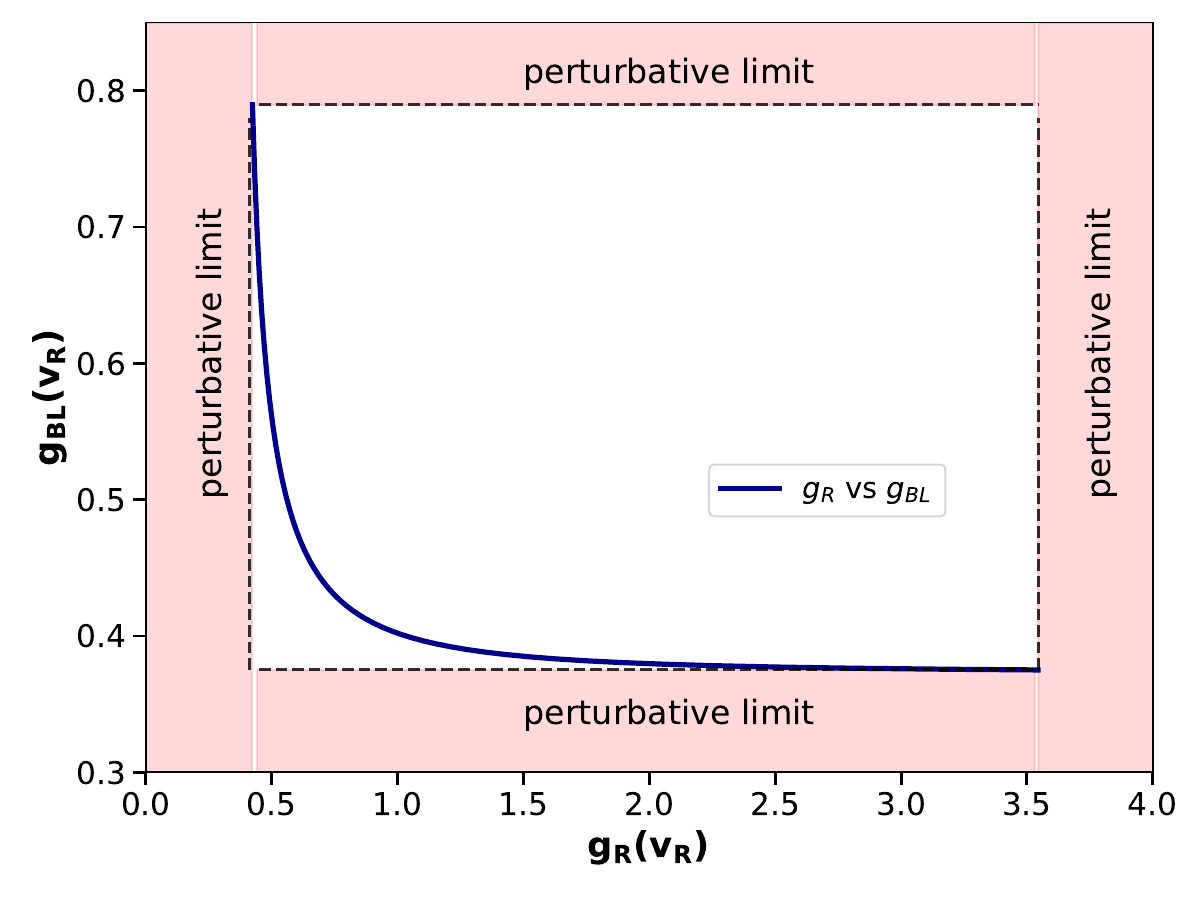}
        \caption{}
        \label{fig:plot_c}
    \end{subfigure}    \caption{\label{corr_gR_gBL}\textit{Left:} Dependence of $g_{BL}$ and $g_R$ on the ratio parameter $r_g= \tfrac{g_R(v_R)}{g_{L}(v_R)}$, shown at $v_R=10$ TeV (solid lines) and at the GUT scale set as $10^{16}$ GeV (dashed lines). \textit{Right:} $g_R$ and $g_{BL}$ at $v_R$. In both panels, the shaded regions are excluded by the perturbative limit ($g_i \le \sqrt{4\pi}$).}
\end{figure}%
The results of our analysis are presented in Fig.~\ref{corr_gR_gBL}. The left panel shows 
the variation of $g_{BL}$ and $g_R$ as functions of $r_g$, evaluated at $v_R$ (solid lines) 
and at the GUT scale (dashed lines). The contrasting behaviour of the two couplings is 
clearly visible: since $g_R$ is asymptotically free, the dashed red line lies below the 
corresponding solid line, whereas $g_{BL}$, which grows with energy, has its dashed blue 
line lying above the solid one. The perturbative limit is approached by $g_{BL}$ at the GUT 
scale for small values of $r_g \sim 0.677$, whilst $g_R(v_R)$ saturates the perturbative 
bound for large values $r_g \sim 5.66$. Furthermore, $g_R(v_R)$ increases with $r_g$ 
whilst $g_{BL}(v_R)$ decreases, reflecting the inverse relationship between the two 
couplings imposed by the hypercharge matching condition of Eq.~\eqref{gygRgBL}. Requiring 
both couplings to remain perturbative at all scales between $v_R$ and the GUT scale yields 
the following allowed ranges:
\begin{equation}
    0.423 < g_R(v
    _R)< \sqrt{4\pi}, \qquad 0.375 < g_{BL}(v_R) < 0.789, \qquad{\rm and }\qquad 0.677 < r_g < 5.660.
\end{equation}
The right panel of Fig.~\ref{corr_gR_gBL} displays the corresponding correlation between  $g_R$ and $g_{BL}$ at the scale $v_R$, with the excluded regions shaded. The shape of the allowed band is a direct consequence of the matching condition 
Eq.~\eqref{gygRgBL}, which constrains $g_R$ and $g_{BL}$ to lie on a curve of fixed 
$1/g_R^2 + 1/g_{BL}^2 = 1/g_Y^2$, so that an increase in $g_R$ necessarily entails 
a decrease in $g_{BL}$ and vice versa.

To assess the sensitivity of these results to the choice of symmetry breaking scale, we 
repeat the analysis for representative values of $v_R$ ranging from 5~TeV to 100~TeV. The 
resulting allowed ranges of $r_g$, $g_R$, and $g_{BL}$ are summarized in 
Table~\ref{gLgBL_constraints}. As $v_R$ increases, the allowed band of $r_g$ shifts 
slightly upward, reflecting the slow logarithmic running of $g_L$ and $g_Y$ between the 
electroweak and symmetry breaking scales. Nevertheless, the overall picture remains 
qualitatively unchanged across this range, and the allowed intervals for $g_R$ and $g_{BL}$ 
are only mildly sensitive to the precise value of $v_R$.

\begin{table}[h!]
\centering
\begin{tabular}{c|c|c|c}
\hline\hline
${v_R}~[\mathrm{TeV}]$ & $\displaystyle{r_g=\frac{g_R}{g_L}}$ & ${g_R}$ & ${g_{BL}}$ \\
\hline
 5   & $[0.665,\, 5.635]$ & $[0.422,\, \sqrt{4\pi}]$ & $[0.370,\, 0.788]$ \\
10   & $[0.677,\, 5.660]$ & $[0.423,\, \sqrt{4\pi}]$ & $[0.375,\, 0.789]$ \\
50   & $[0.695,\, 5.730]$ & $[0.430,\, \sqrt{4\pi}]$ & $[0.382,\, 0.812]$ \\
100  & $[0.702,\, 5.761]$ & $[0.432,\, \sqrt{4\pi}]$ & $[0.385,\, 0.824]$ \\
\hline\hline
\end{tabular}
\caption{Allowed ranges of $r_g$, $g_R$, and $g_{BL}$ at the scale $v_R$ for 
the representative values of the ${SU}(2)_{R'}$ breaking scale, derived by requiring 
both couplings to remain perturbative up to the GUT scale.}
\label{gLgBL_constraints}
\end{table}

\begin{figure}[t]
\centering
\includegraphics[width=.48\textwidth]{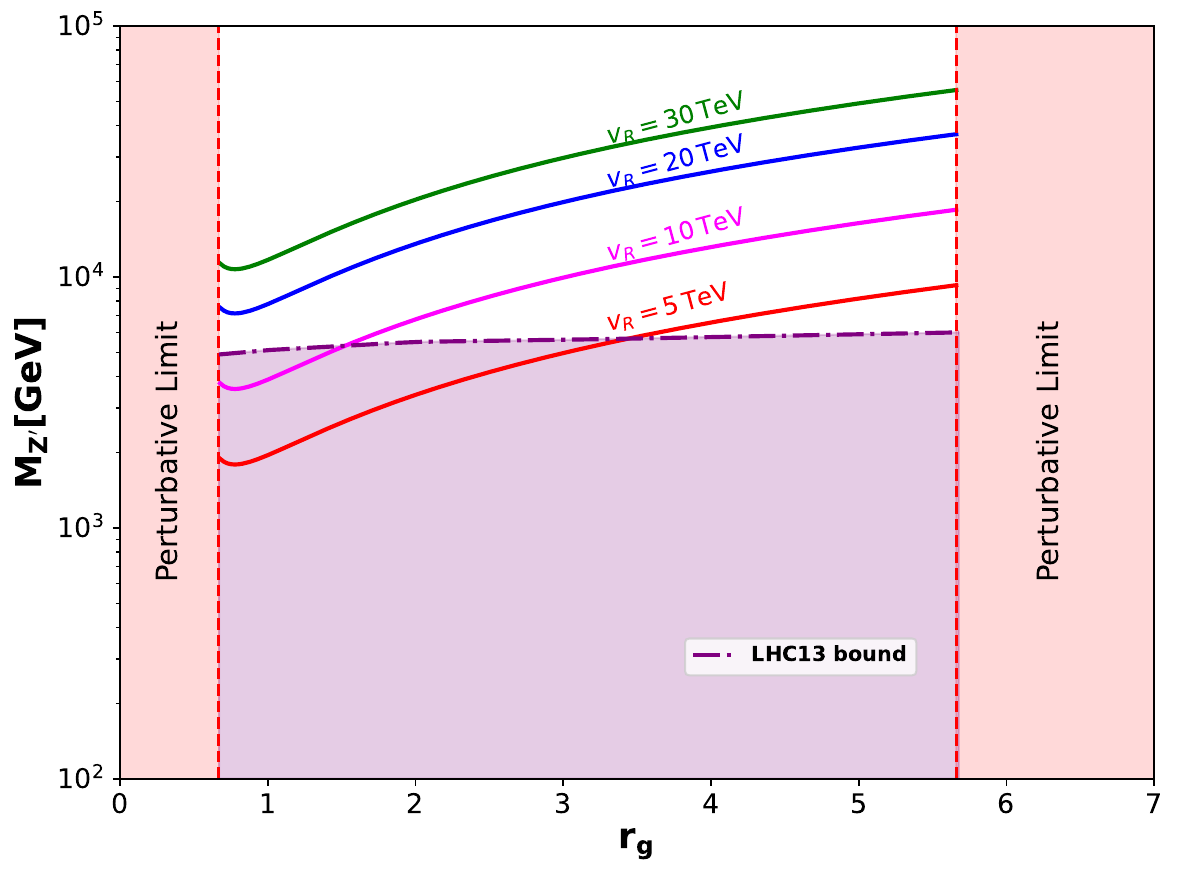}
\hfill
\includegraphics[width=.48\textwidth]{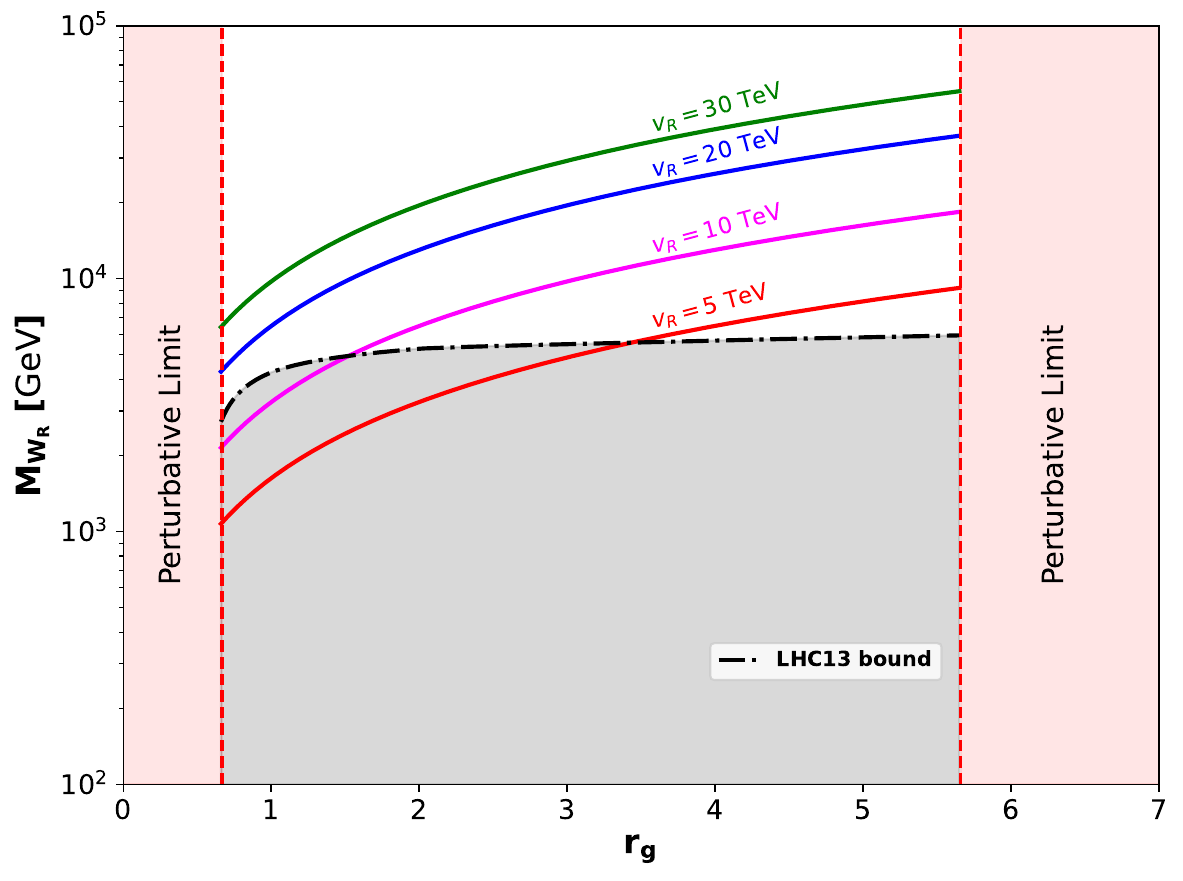}
\caption{The perturbativity bounds on the masses of the heavy gauge bosons $Z'$ (left) and 
$W_R$ (right) as functions of $r_g$, shown for the representative values of $v_R$. The red 
shaded regions are excluded by the perturbativity constraints of 
Table~\ref{gLgBL_constraints}, and the purple dot-dashed line indicates the rescaled LHC 
exclusion limit from $Z' \to \ell^+\ell^-$ searches~\cite{CMS:2021ctt, CMS:2016abv}. The dot-dashed  curve in the right panel shows the indirect lower 
bound on $M_{W_R}$  derived from the $M_{Z'}$ constraint via 
Eqs.~\eqref{eq:MWRmass}--\eqref{eq:MZmass}.}
\label{WR_ZR}
\end{figure}
In Fig.~\ref{WR_ZR}, we present the perturbativity bounds on the masses of the heavy  gauge bosons $Z'$ (left panel) and $W_R$ (right panel) as functions of $r_g$, shown for  several representative values of $v_R$. The $Z'$ mass receives contributions from both  $g_{BL}$ and $g_R$, whilst the $W_R$ mass depends solely on $g_R$, as it arises entirely from the non-abelian ${SU}(2)_{R'}$ sector. The red shaded regions in each panel indicate the parameter space excluded by the perturbativity constraints on $r_g$ summarized in Table~\ref{gLgBL_constraints}.

The current exclusion limit from 13~TeV LHC searches for $Z' \to \ell^+\ell^-$ 
$(\ell = e, \mu)$ is $M_{Z'} \gtrsim 5.1$~TeV~\cite{CMS:2021ctt, CMS:2016abv}, derived under the assumption $g_R = g_L$ (i.e.\ $r_g = 1$). 
For $g_R \neq g_L$, this limit must be rescaled, since the $Z'$ production cross section  $\sigma(pp \to Z' \to \ell^+\ell^-)$ depends on $r_g$. After applying this rescaling, the excluded $Z'$ mass range shifts to approximately $[4.9,\, 6.1]$~TeV for $r_g \in [0.677,\, 5.66]$, as indicated by the purple dot-dashed line in the left panel of Fig.~\ref{WR_ZR}. This translates into a lower bound on $r_g$ that depends on $v_R$; for instance, for $v_R = 10$~TeV the constraint requires $r_g \gtrsim 1.1$.

Since the $W_R$  boson decays exclusively into a SM up-type quark and an exotic down-type 
quark, or into a charged lepton and a scotino, it does not give rise to the purely SM final states and therefore evades the direct LHC search constraints entirely. However, the $W_R$ and $Z'$ masses are related through Eqs.~\eqref{eq:MWRmass} and~\eqref{eq:MZmass}, so that the lower bound on $M_{Z'}$ translates into an indirect lower bound on $M_{W_R}$, as discussed in Ref.~\cite{Frank:2024imi}. This indirect limit is shown as the dot-dashed curve in the right panel of Fig.~\ref{WR_ZR}, and is found to be numerically very close to the direct $r_g$ constraint derived from $M_{Z'}$. In principle, the mass ratio $M_{W_R}/M_{Z'}$ 
carries a mild dependence on $\tan\beta$ through the gauge boson mass formulae; however, we have verified explicitly that this dependence is numerically negligible over the parameter space considered here and can safely be ignored.

\subsection{ Scalar Quartic Couplings}

We now turn to the one-loop RGE running of the scalar quartic couplings, whose explicit 
$\beta$-functions are collected in Appendix~\ref{app:RGE}. The running of these couplings 
is driven by the contributions from the quartic sector itself, as well as from the gauge and 
the Yukawa sectors.

To illustrate the qualitative features of this evolution, we select three benchmark points 
(BPs) from the parameter scan of Sec.~\ref{sec:theroeticalscan}, each satisfying the 
combined constraints of the vacuum stability, the perturbative unitarity, and the perturbativity, with 
$v_R = 10$~TeV fixed. The benchmark values of the independent quartic couplings, 
$\tan\beta$, and the trilinear coupling $\mu_3$ are given in Table~\ref{tab:BPsRGE}. The 
Yukawa couplings entering the $\beta$-functions are fixed by the fermion mass spectrum; we 
take scotino masses $(m_{n_e},\, m_{n_\mu},\, m_{n_\tau}) = (300,\, 400,\, 500)$~GeV, 
exotic quark masses $(m_{d'},\, m_{s'},\, m_{b'}) = (1.5,\, 2,\, 3)$~TeV, and the 
remaining standard fermion masses at their measured values. The gauge couplings are fixed 
with $r_g = 1$ for simplicity, as discussed in Sec.~\ref{subsec:gaugecouplings}.

\begin{table}[ht]
\centering
\begin{tabular}{l|r|r|r|r|r|r}
\hline\hline
BP & $\alpha_{12}$ & $\alpha_{13}$ & $\lambda_{2}$ & $\lambda_{3}$ & 
$\tan\beta$ & $\mu_3$~[GeV] \\
\hline
BP1 & $0.0272$ & $0.0270$ & $-0.1531$ & $0.00425$ & $15.41$ & $-333.69$ \\
BP2 & $0.0627$ & $0.0625$ & $-0.1495$ & $0.00461$ & $15.83$ & $-343.92$ \\
BP3 & $0.0182$ & $0.0180$ & $-0.1047$ & $0.00425$ & $17.81$ & $-333.67$ \\
\hline\hline
\end{tabular}
\caption{Input values of the scalar potential parameters at the scale $v_R = 10$~TeV for 
the three benchmark points used to illustrate the RGE running of the quartic couplings in 
Fig.~\ref{fig:bp_running}.}
\label{tab:BPsRGE}
\end{table}

\begin{figure}[t]
\centering
\includegraphics[width=0.48\textwidth]{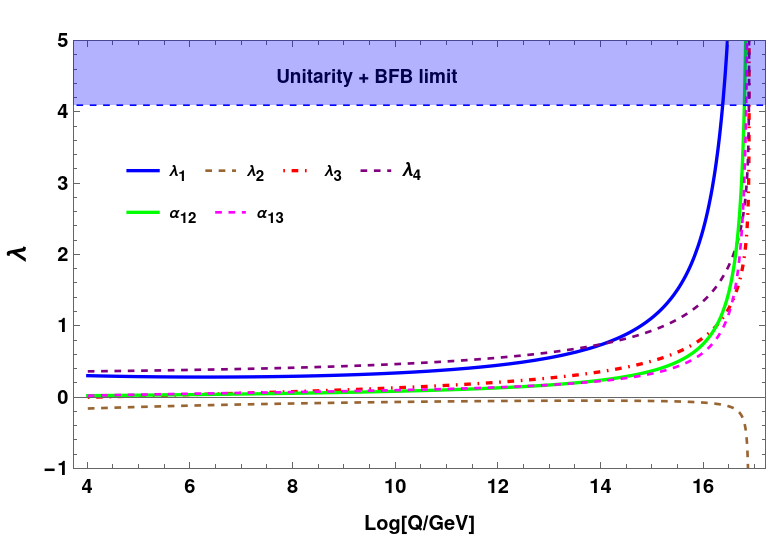} \hfill
\includegraphics[width=0.48\textwidth]{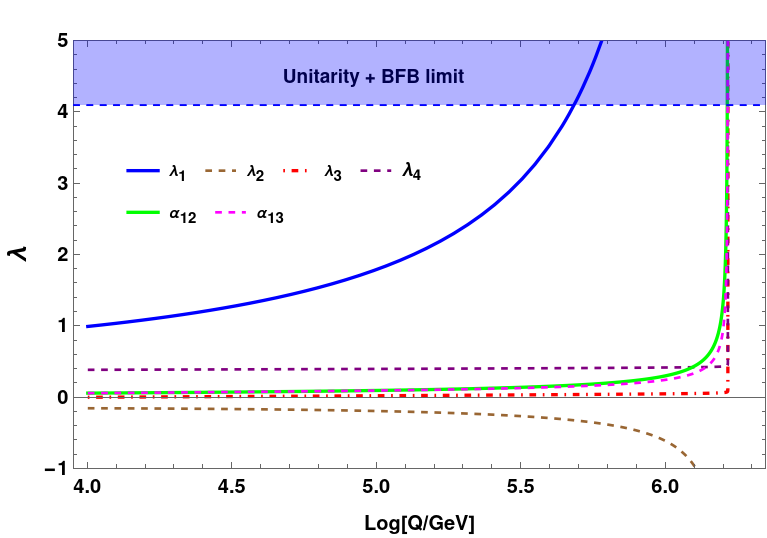}
\vspace{2mm}
\includegraphics[width=0.60\textwidth]{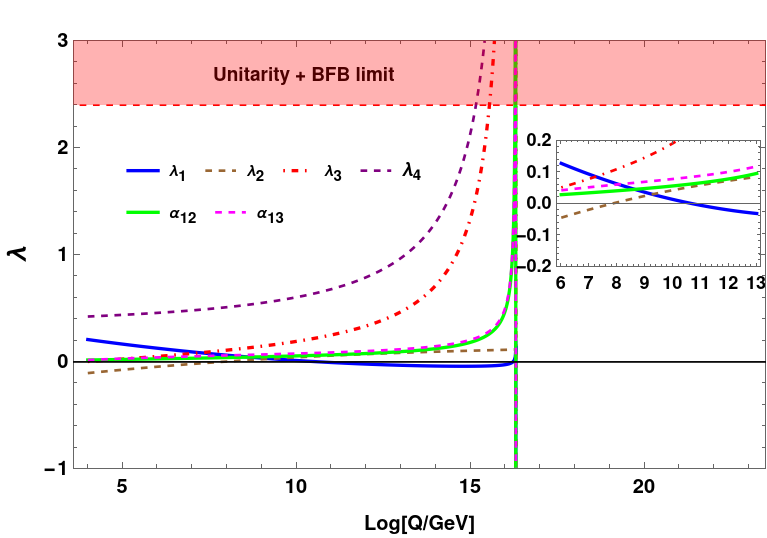}

\caption{RGE running of the scalar quartic couplings from the input scale $Q = v_R = 
10$~TeV for BP1 (top left), BP2 (top right), and BP3 (bottom), as defined in 
Table~\ref{tab:BPsRGE}. The purple shaded band indicates the theoretical upper bound 
on $\lambda_1$, and the red shaded band indicates the theoretical bound on 
$\lambda_3$, both derived from the combined vacuum stability and perturbative unitarity 
conditions of Eq.~\eqref{eq:ThConstraints}. Couplings must remain below the upper band 
and above the lower band to satisfy all theoretical constraints.}
\label{fig:bp_running}
\end{figure}

The resulting RGE evolution of the quartic couplings is shown in Fig.~\ref{fig:bp_running} 
for each of the three benchmark points. The shaded horizontal bands indicate the upper 
limits imposed by the combined vacuum stability and the perturbative unitarity constraints of 
Eq.~\eqref{eq:ThConstraints}, with values inside the bands being excluded; for clarity, 
only the most constraining bound is displayed for each coupling, as the remaining 
constraints are satisfied up to higher scales and do not limit the cutoff scale in the cases shown.
The three benchmark points illustrate qualitatively distinct patterns of RGE evolution, which we now discuss in turn.
The benchmark BP1 (Fig.~\ref{fig:bp_running}~{\it top left}) remains theoretically consistent up to approximately 
$10^{16}$~GeV, at which point $\lambda_1$ reaches the upper boundary imposed by the 
combined vacuum stability and the unitarity constraints. The Landau pole lies roughly an order of magnitude above this scale, so it is the violation of the theoretical bounds rather than a divergence that sets the cutoff.
In the case of BP2 (Fig.~\ref{fig:bp_running}~{\it top right}), the parameters $\lambda_2$, $\lambda_3$, and $\tan\beta$ are nearly unchanged relative to BP1, whilst $\alpha_{12}$ and $\alpha_{13}$ are approximately twice as large at the input scale. Through the SM Higgs mass condition that fixes $\lambda_1$, this results in a substantially larger initial value of $\lambda_1$, as it is evident from the figure. This larger initial value drives a much faster running of $\lambda_1$, which violates the theoretical constraints already at around $10^{5.5}$~GeV, with the Landau pole reached shortly after at $10^6$~GeV. The accelerated running of $\lambda_1$ in turn feeds into $\lambda_2$ through the $24\lambda_1\lambda_2$ term in $\beta(\lambda_2)$, causing $\lambda_2$ to also run faster than in BP1. In contrast, $\lambda_3$ and $\lambda_4$ are insensitive to $\lambda_1$ and $\lambda_2$ through their $\beta$-functions, and consequently follow a trajectory very similar to BP1. On the other hand,  in the case of BP3 (Fig.~\ref{fig:bp_running}~{\it bottom }), the Landau pole lies slightly above $10^{16}$~GeV, yet the 
theoretical constraints are violated considerably earlier by two distinct couplings. First, $\lambda_1$ turns negative around $10^{10}$~GeV, violating the BFB condition, then $\lambda_2$ crosses zero and becomes positive at around $10^8$~GeV, violating the 
copositivity condition required for a stable vacuum. The remaining couplings follow a 
qualitatively similar trajectory to BP1, with $\lambda_3$ and $\lambda_4$ approaching the 
unitarity bound near $10^{15}$~GeV.


The benchmark analysis above serves to illustrate the high sensitivity of the RGE 
evolution to the input values of the couplings at the scale $v_R$, and motivates a 
systematic study of the energy range over which the model remains theoretically 
consistent. We now perform such an analysis, scanning the input parameter space for 
$v_R \in \{5,\, 10,\, 15,\, 20\}$~TeV and determining, for each point, the highest scale 
up to which all theoretical constraints are simultaneously satisfied.

Specifically, for a given $v_R$, we perform a scan over the initial parameter space as 
defined in Eq.~\eqref{uni_scan}, retaining only those points that satisfy the vacuum 
stability conditions of Eq.~\eqref{BFB_conditions} at the input scale. In the subsequent RGE 
evolution, we require that all quartic couplings remain within their perturbative limits at every scale, with the additional sign conditions that all couplings except $\lambda_2$ remain non-negative, whilst $\lambda_2$ must remain strictly negative, as dictated by the copositivity and the vacuum conditions throughout the running. For each accepted point, the couplings are evolved from $v_R$ upward, and their values are recorded at a series of the cutoff scales ranging from $10^5$~GeV to $10^{16}$~GeV. The cutoff scale assigned to each point is the lowest scale at which any one of the following conditions is first violated: the vacuum stability bounds, the perturbative unitarity constraints, the sign conditions on the couplings, or the perturbativity limit. In this way, we map out the region of input parameter space that remains theoretically reliable up to any given energy scale, and identify how this region shrinks as the target cutoff scale is increased.

A parameter point is retained as theoretically viable up to a chosen cutoff scale if it satisfies all of the following conditions throughout the RGE evolution from $v_R$ up to that scale: (1) no Landau pole is encountered below the cutoff scale, (2) the vacuum stability (BFB) conditions of Eq.~\eqref{BFB_conditions} are satisfied 
at every scale  up to the cut off scale, and
(3) all couplings remain within their perturbative limits and within the bounds imposed by the perturbative unitarity constraints of Eqs.~\eqref{eq:eigen1n2}--\eqref{eq:eigen13n14} at every scale.
The set of input values passing all three criteria defines the allowed initial parameter space consistent with theoretical reliability up to the chosen cutoff scale.

\begin{figure}[h]
\centering
\includegraphics[width=0.48\textwidth]{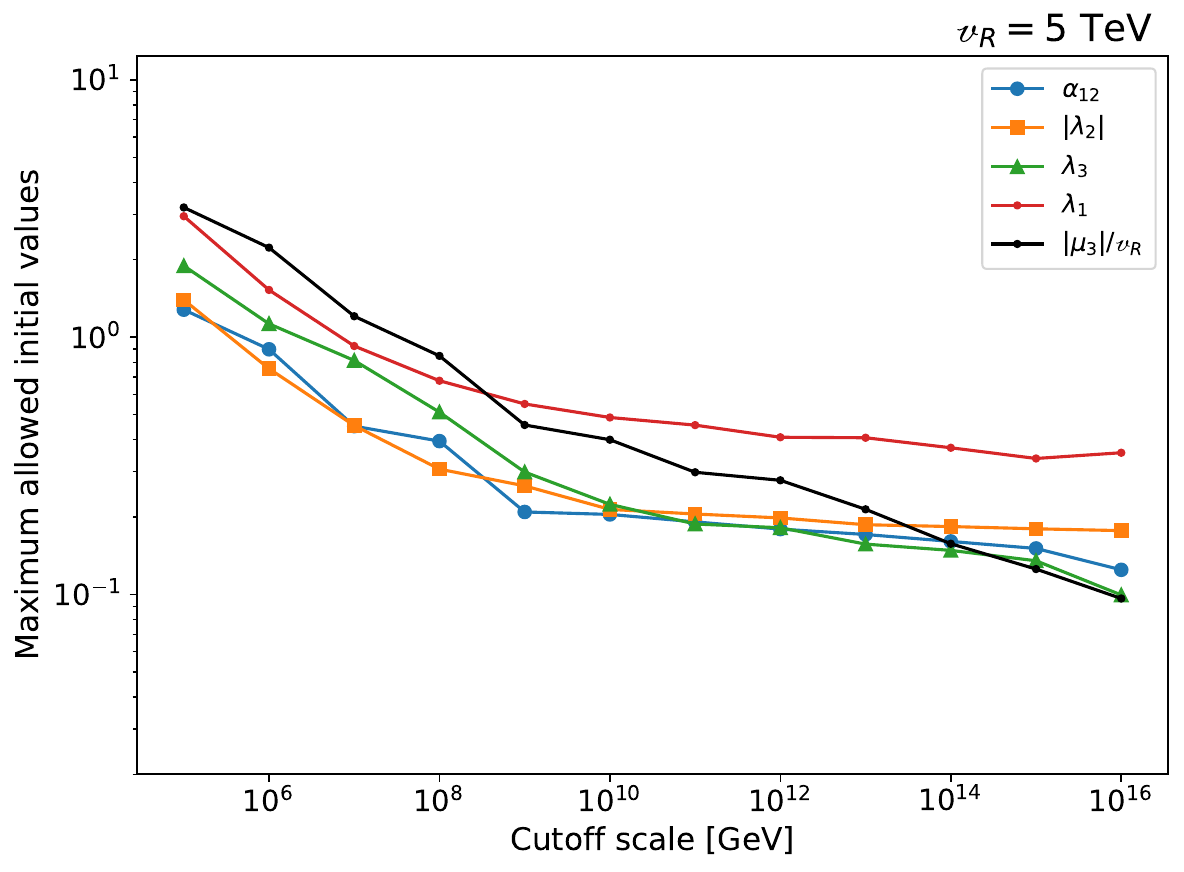}\hfill
\includegraphics[width=0.48\textwidth]{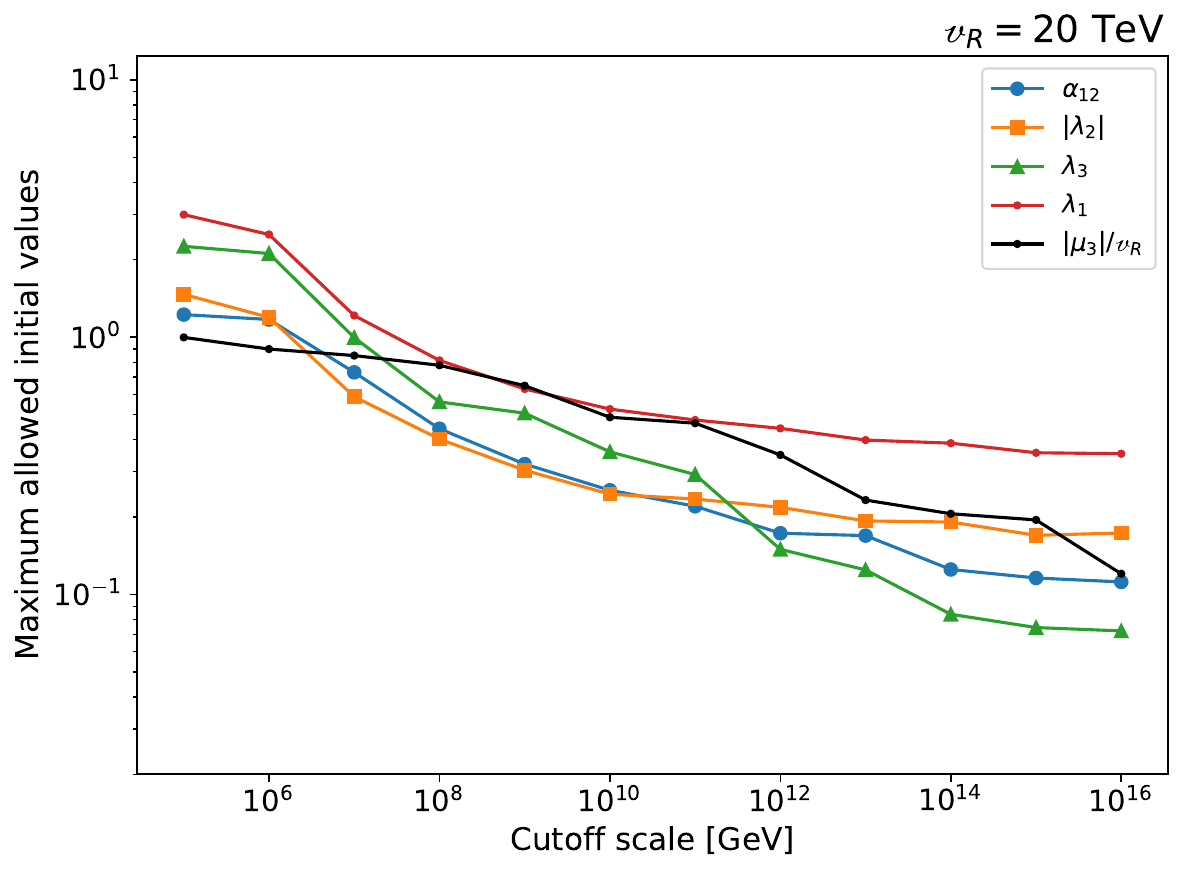}
\caption{The maximum allowed initial values of the scalar quartic couplings 
$\alpha_{12} \simeq \alpha_{13}$ (blue), $|\lambda_2|$ (orange), $\lambda_3$ (green), 
$\lambda_1$ (red), and $|\mu_3|/v_R$ (black), as functions of the cutoff scale, for 
$v_R = 5$~TeV (left) and $v_R = 20$~TeV (right). Each curve represents the upper boundary 
of the theoretically viable parameter space at the corresponding cutoff scale, obtained by 
requiring simultaneously the absence of Landau poles, the satisfaction of BFB conditions, 
and compliance with the perturbative unitarity bounds throughout the RGE evolution.}
\label{contraint_limit}
\end{figure}

For each cutoff scale, we extract the maximum allowed initial value of each coupling consistent with these criteria. The results are presented in Fig.~\ref{contraint_limit}, 
which shows the upper limits on $\alpha_{12} \simeq \alpha_{13}$ (blue), $|\lambda_2|$ 
(orange), $\lambda_3$ (green), $\lambda_1$ (red), and $|\mu_3|/v_R$ (black) as functions of the cutoff scale. Since the difference between $\alpha_{12}$ and $\alpha_{13}$ is negligible over the allowed parameter space, we set $\alpha_{12} = \alpha_{13}$ in the presentation. The results are shown for $v_R = 5$~TeV (left panel) and $v_R = 20$~TeV (right panel) to illustrate the sensitivity to the symmetry breaking scale; both panels are found to be in close agreement, confirming that the theoretical bounds on the quartic couplings are only mildly sensitive to the precise value of $v_R$.

As expected on the general grounds, the maximum allowed values of all couplings decrease 
monotonically as the cutoff scale is raised, reflecting the fact that a higher cutoff 
imposes more stringent requirements on the RGE trajectory. This behaviour has direct 
phenomenological consequences: the choice of cutoff scale translates into concrete upper 
bounds on the scalar couplings at the electroweak scale, which in turn constrain the physical Higgs spectrum and the strength of scalar interactions accessible at the collider experiments. The interplay between the target validity scale and the low-energy phenomenology of the model is therefore an important consideration in any phenomenological study of the ALRM.

\subsection{Constraints on scalar masses:}
The masses of the physical scalars in the ALRM are determined by the couplings of the 
scalar potential and the vacuum expectation values, so the theoretical constraints derived 
above translate directly into upper bounds on the scalar mass spectrum. From 
Eqs.~\eqref{eq:mH10A1}, \eqref{eq:mH1p}, and~\eqref{eq:mH2p}, one can see that for a 
fixed $v_R$, the scalar masses are governed primarily by $\mu_3$ and $\tan\beta$, with 
subleading corrections from the quartic couplings. In particular, the mass splitting 
between $H_2^\pm$ and $H_1^0/A_1$ is controlled by $\lambda_2$ and the VEV $v_u$; this 
splitting reaches at the most around 50~GeV when the cutoff scale is set to $10^{16}$~GeV, 
and grows to roughly an order of magnitude larger for the cutoff scale at $10^5$~GeV.

The dominant contributions to the charged Higgs masses take the approximate forms as 
\begin{equation}
m^2_{H_1^\pm} \approx -\frac{\mu_3 v_R}{\sqrt{2}}\tan\beta, \qquad {\rm and} \qquad
m^2_{H_2^\pm} \approx -\frac{\mu_3 v_R}{\sqrt{2}}\cot\beta,
\end{equation}
where $\mu_3 < 0$ ensures the positive mass-squared eigenvalues. However, the vacuum 
stability and the perturbative unitarity constraints couple $\tan\beta$ and $\mu_3$; the 
larger values of $|\mu_3|$ restricting the maximum allowed $\tan\beta$ to approximately 
8, as visible in Fig.~\ref{BFB_Unitarity}. Ignoring the RGE constraints and working 
within the tree-level allowed parameter space alone, one would naively estimate upper 
bounds of
\begin{equation}
m_{H_1^\pm} \sim 20~\text{TeV} \qquad \text{and} \qquad m_{H_2^\pm} \sim 4~\text{TeV}.
\end{equation}
\begin{figure}[h]
\centering
\includegraphics[width=0.45\textwidth]{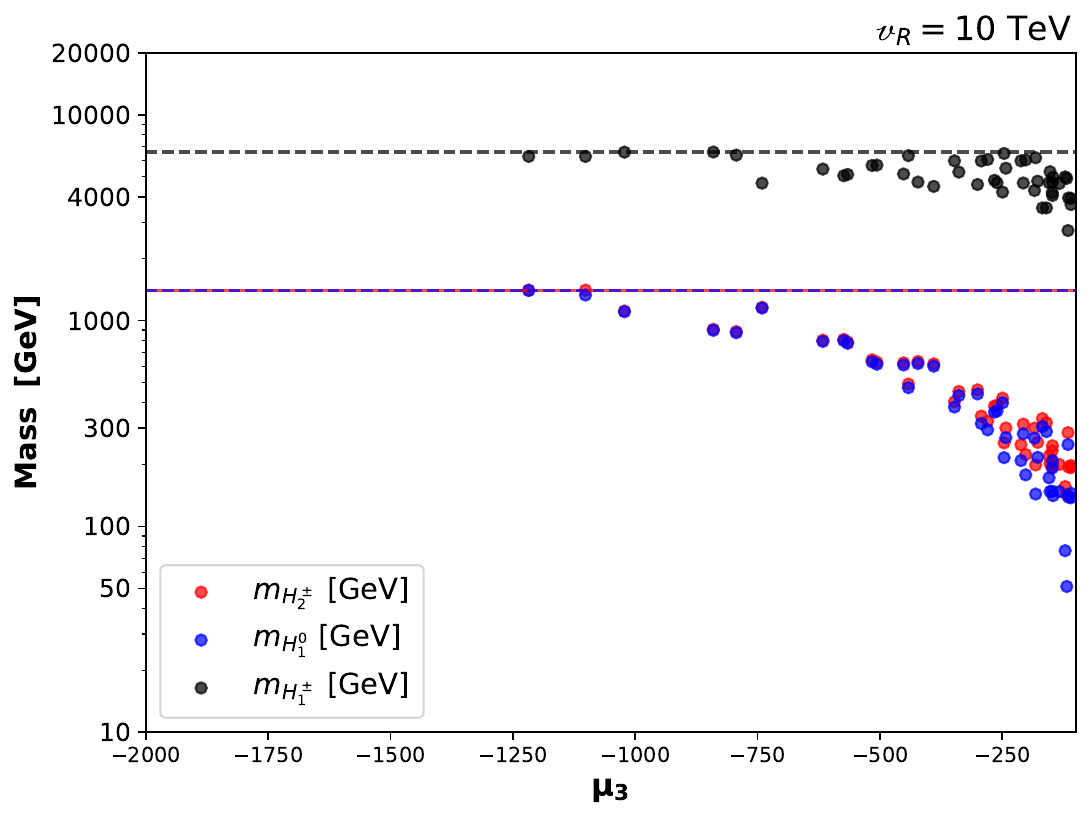}\hfill
\includegraphics[width=0.43\textwidth]{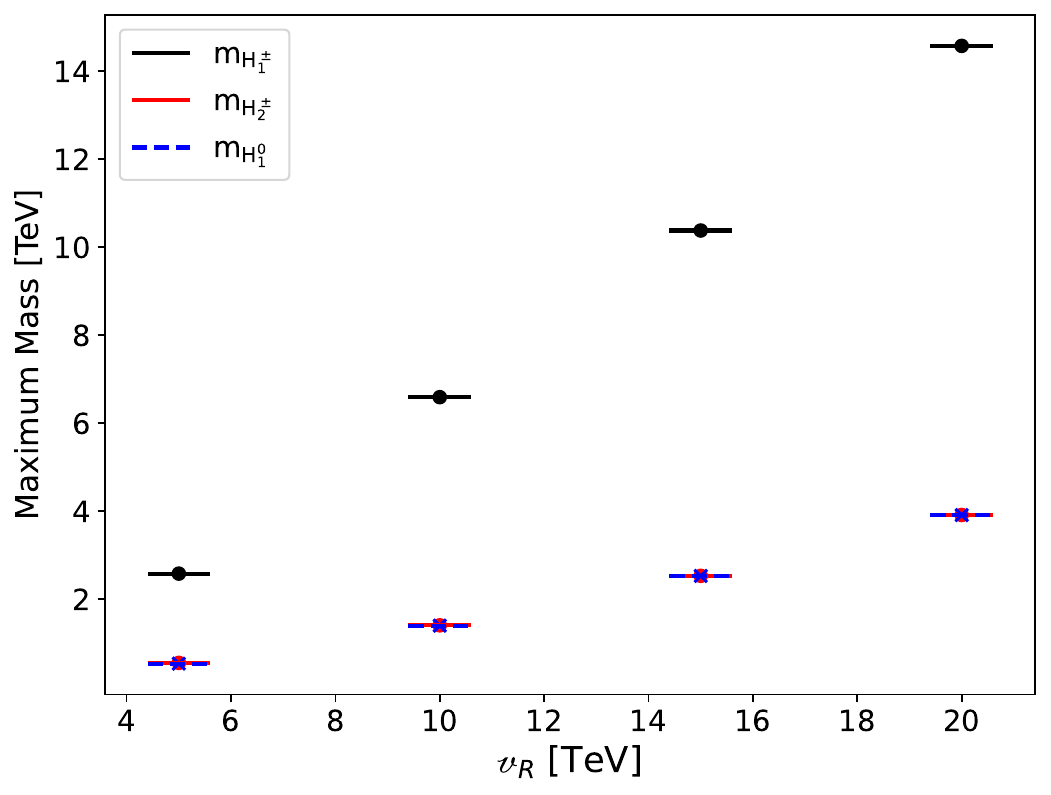}
\caption{Left: Mass spectrum of $m_{H_2^\pm}$ (red), $m_{H_1^0}$ (blue), and $m_{H_1^\pm}$ (black) as a function of $\mu_3$ for $v_R = 10$~TeV, obtained from the full numerical scan with all theoretical constraints imposed up to $10^{16}$~GeV. Right: Maximum allowed scalar masses as a function of $v_R$, derived by requiring theoretical consistency up to $10^{16}$~GeV.}
\label{mass_const}
\end{figure}
The RGE analysis, however, significantly tightens these estimates. Requiring the model to remain theoretically consistent up to $10^{16}$~GeV, it restricts $|\mu_3|$ to 
approximately 1.3~TeV for $v_R = 10$~TeV, which reduces 
the maximum allowed masses to
\begin{equation}
m_{H_1^\pm} \sim 6.5~\text{TeV} \qquad \text{and} \qquad m_{H_2^\pm} \sim 1.5~\text{TeV}.
\end{equation}
These analytical estimates are confirmed by the full numerical scan, whose results for 
$v_R = 10$~TeV are shown in the left panel of Fig.~\ref{mass_const}. The light charged 
Higgs $H_2^\pm$ (red points) reaches a maximum mass of approximately 1.5~TeV, while the degenerate neutral states $H_1^0$ (blue points) and the heavy charged Higgs $H_1^\pm$ 
(black points) are bounded by approximately 1.3~TeV and 6.5~TeV, respectively. The 
dependence of these upper bounds on $v_R$ is shown in the right panel of Fig.~\ref{mass_const}: larger values of $v_R$ push all scalar masses upward, with $H_2^\pm$ entering the multi-TeV range whilst $H_1^\pm$ and $H_1^0$ become correspondingly heavier.

Taken together, these results demonstrate that the combination of the vacuum stability conditions and the perturbative unitarity constraints, when consistently imposed through the RGE, places stringent and calculable upper bounds on the scalar mass spectrum of the ALRM. A key feature of these bounds is that they are not merely order-of-magnitude estimates but are genuinely predictive: for a given choice of $v_R$ and a specified validity scale, the maximum allowed masses of $H_2^\pm$, $H_1^0$, and $H_1^\pm$ are concretely determined, as summarised in Fig.~\ref{mass_const}.
These bounds are therefore of direct relevance for ongoing and future searches for 
extended Higgs sectors at the LHC and at the future high-energy colliders. We emphasise, 
however, that the numerical values of the upper bounds carry a dependence on $v_R$: 
larger symmetry breaking scales allow heavier scalar states, as illustrated in the right 
panel of Fig.~\ref{mass_const}, so that a complete phenomenological assessment requires 
specifying both the validity scale and the value of $v_R$.
 
\section{Summary and Conclusions}\label{sec:conclusion}

We have performed a systematic study of the theoretical consistency of the Alternative 
Left-Right Model (ALRM), an $E_6$-motivated extension of the Standard Model that 
naturally accommodates neutrino masses and dark matter whilst avoiding tree-level 
flavour-changing neutral currents. Three classes of theoretical constraints have been imposed on the scalar sector: the perturbative unitarity, the vacuum stability, and the
perturbativity. We derived for the first time the complete set of perturbative unitarity constraints on 
the ALRM scalar potential, obtaining 14 independent conditions on the quartic couplings, verified independently using \texttt{SARAH} and \texttt{anyPUB}. Combined with the boundedness-from-below conditions and the requirement of positive-definite scalar mass-squared eigenvalues, these constraints are found to be complementary: each excludes regions of parameter space not ruled out by the other, and their simultaneous imposition yields significantly more stringent restrictions than by either set alone. The resulting bounds on the quartic couplings are $(\alpha_{12},\,\alpha_{13}) \lesssim 1.4$, $\lambda_1 \lesssim 3.0$, $|\lambda_2| \lesssim 1.5$, $\lambda_3 \lesssim 2.5$, and $\lambda_4 \lesssim 2\pi$. 

We then performed a one-loop RGE analysis of all model parameters, implemented in two 
stages reflecting the two-step symmetry breaking of the ALRM. The RGE running reveals 
that the tree-level allowed parameter space is further and substantially restricted when 
theoretical consistency is demanded up to higher energy scales. The allowed ranges of 
all quartic couplings shrink monotonically as the target validity scale is raised, with 
the constraints becoming particularly stringent when the model is required to remain well-behaved up to a high energy cut-off scale. The analysis also exposes the structure underlying this running: the cutoff scale is highly sensitive to the initial value of $\lambda_1$, which is fixed by the SM Higgs mass condition and inherits a strong dependence on 
$\alpha_{12}$ and $\alpha_{13}$, whilst $\lambda_3$ and $\lambda_4$ are comparatively 
insensitive owing to the structure of their $\beta$-functions. For the gauge sector, 
perturbativity up to a scale of $10^{16}$ GeV constrains the ratio $r_g = g_R/g_L$ to lie in the 
range $0.677 \lesssim r_g \lesssim 5.660$ for $v_R = 10$~TeV, with only mild dependence 
on $v_R$ across the range of $5$ TeV to $100$~TeV.
As a consequence of the restricted parameter space, the physical scalar masses of the 
ALRM are also bounded from above. For $v_R = 10$~TeV, requiring theoretical consistency 
up to a cut-off scale of $10^{16}$ GeV yields
\begin{equation}
m_{H_1^\pm} \lesssim 6.5~\text{TeV}, \qquad 
m_{H_2^\pm} \lesssim 1.5~\text{TeV}, \qquad 
m_{H_1^0} \simeq m_{A_1} \lesssim 1.3~\text{TeV},
\end{equation}
with all bounds scaling with $v_R$. These upper limits provide a predictive window for 
searches for the extended Higgs sector of the ALRM at the LHC and the future collider experiments, and 
allow the model to be systematically constrained or excluded when no such states be observed within the expected mass ranges.

In conclusion, our analysis demonstrates that the consistent combination of the vacuum 
stability, the perturbative unitarity, and one-loop RG evolution provides a powerful 
framework for constraining the ALRM, restricting the parameter space far more 
stringently than tree-level considerations alone. These results provide a 
systematic theoretical foundation for future phenomenological studies of the model, 
including collider searches for its exotic scalar, gauge boson, and fermionic states.

\section{Acknowledgements}
HD thanks to the Council of Scientific \& Industrial Research (CSIR), Govt. of India for the senior research fellowship. The work of SKG has been supported by SERB, DST, India through grant TAR/2023/000116. SKG acknowledges the Manipal Centre for Natural Sciences, Centre of Excellence, Manipal Academy of Higher Education (MAHE) for facilities and support. P. P. and A. thank the ANRF, India for financial support with SERB research grant (CRG/2022/002670).

\appendix
\section{Tadpole Equations}\label{tadpole}
With $\langle \phi_2^{0R}\rangle = v_u,~\langle \chi_L^{0R}\rangle = v_L,~\langle \chi_R^{0R}\rangle = v_R$, where the superscript $R$ denotes the real part of the corresponding field, and $\langle \psi\rangle = 0$ for all other field components, the nontrivial minimization conditions of the potential are  given by
\begin{align} 
\left.\frac{\partial V}{\partial \phi_{2}^{0R}}\right|_{\rm vacuum} =0~\implies~~& \frac{1}{\sqrt{2}} \mu_3 v_L v_R  + \lambda_1 v_{u}^{3}  + v_u \Big(   \alpha_{12}\Big(v_L^2+v_R^2 \Big) - \mu^2_1 \Big)=0\\ 
\left.\frac{\partial V}{\partial \chi_{L}^{0R}}\right|_{\rm vacuum} =0~\implies~~&= \frac{1}{\sqrt{2}} \mu_3 v_u v_R  + \lambda_{3} v_{L}^{3}  + v_L \Big(\alpha_{12} v_{u}^{2}  + \lambda_4 v_{R}^{2}  - \mu^2_{2} \Big) =0\\ 
\left.\frac{\partial V}{\partial \chi_{R}^{0R}}\right|_{\rm vacuum} =0~\implies~~&= \frac{1}{\sqrt{2}} \mu_3 v_u v_L  + v_R \Big(\alpha_{12}v_{u}^{2}  + \lambda_{3} v_{R}^{2}  + \lambda_4 v_{L}^{2}  - \mu^2_{2} \Big)=0.
\end{align} 
Solving these, we can express the coefficient of the quadratic  terms, $\mu_1^2$ and $\mu_2^2$, and one of the quartic couplings $\lambda_4$ as
\begin{align}\label{lam4_lam3_rel}
    \mu_{1}^2 &= 
        \lambda_1 v_u^2 +\alpha_{12}(v_L^2+v_R^2)+ \frac{\mu_3 v_L v_R}{\sqrt{2}v_u} \nonumber
    , \\
    \mu_{2}^2 &= (v_L^2+v_R^2)\lambda_3+ \alpha_{12}v_u^2, \nonumber
   \\
    \lambda_4& = \lambda_3-\frac{\mu_3 v_u}{\sqrt{2}v_L v_R}.
    \end{align}


\section{Renormalization Group Equations}\label{app:RGE}
With the convention
\begin{equation*}
\beta\left(X\right) \equiv \mu \frac{d X}{d \mu}\equiv \frac{1}{\left(4 \pi\right)^{2}}\beta^{(1)}(X)
\end{equation*}
the 1-loop beta functions of the various parameters are given by\footnote{These equations are presented in Ref. \cite{Deka:2026lrh}.}
\begin{align}
&\beta^{(1)}(g_{BL}) =\frac{13}{3} g_{BL}^{3}\\
&\beta^{(1)}(g_L) =- \frac{17}{6} g_L^{3}\\
&\beta^{(1)}(g_R) =- \frac{17}{6} g_R^{3}\\
&\beta^{(1)}(g_3) =-5 g_3^{3}
\end{align}

\begin{align*}
\beta^{(1)}(Y_{Q1}) =& ~Y_{Q2} Y_{Q2}^{\dagger} Y_{Q1}+ \frac{3}{2} Y_{Q1} Y_{Q1}^{\dagger} Y_{Q1}+ 3 {\text{Tr}}\left(Y_{Q1}^{\dagger} Y_{Q1} \right) Y_{Q1}
+ {\text{Tr}}\left(Y_{L2}^{\dagger} Y_{L2} \right) Y_{Q1}-  \frac{5}{12} g_{BL}^{2} Y_{Q1} \\ &-  \frac{9}{4} g_L^{2} Y_{Q1}- 8 g_3^{2} Y_{Q1}
\end{align*}
\begin{align*}
\beta^{(1)}(Y_{Q2}) = &~ 2 Y_{Q2} Y_{Q2}^{\dagger} Y_{Q2}+ \frac{1}{2} Y_{Q2} Y_{Q3} Y_{Q3}^{\dagger}+ \frac{1}{2} Y_{Q1} Y_{Q1}^{\dagger} Y_{Q2}
+ 3 {\text{Tr}}\left(Y_{Q2}^{\dagger} Y_{Q2} \right) Y_{Q2}+ {\text{Tr}}\left(Y_{L1}^{\dagger} Y_{L1} \right) Y_{Q2}\nonumber \\&  -\frac{1}{6} g_{BL}^{2} Y_{Q2}-  \frac{9}{4} g_L^{2} Y_{Q2}-  \frac{9}{4} g_R^{2} Y_{Q2}- 8 g_3^{2} Y_{Q2}
\end{align*}

\begin{align*}
\beta^{(1)}(Y_{Q3}) = &~Y_{Q2}^{\dagger} Y_{Q2} Y_{Q3}+ \frac{3}{2} Y_{Q3} Y_{Q3}^{\dagger} Y_{Q3}+ 3 {\text{Tr}}\left(Y_{Q3}^{\dagger} Y_{Q3} \right) Y_{Q3}
+ {\text{Tr}}\left(Y_{L3}^{\dagger} Y_{L3} \right) Y_{Q3}-  \frac{5}{12} g_{BL}^{2} Y_{Q3}\\ &-  \frac{9}{4} g_R^{2} Y_{Q3}- 8 g_3^{2} Y_{Q3}
\end{align*}
\begin{align*}
\beta^{(1)}(Y_{L1}) =&~ 2 Y_{L1} Y_{L1}^{\dagger} Y_{L1}+ \frac{1}{2} Y_{L1} Y_{L3} Y_{L3}^{\dagger}+ \frac{1}{2} Y_{L2} Y_{L2}^{\dagger} Y_{L1}+ 3 {\text{Tr}}\left(Y_{Q2}^{\dagger} Y_{Q2} \right) Y_{L1}-  \frac{3}{2} g_{BL}^{2} Y_{L1} -  \frac{9}{4} g_R^{2} Y_{L1}\\ &+ {\text{Tr}}\left(Y_{L1}^{\dagger} Y_{L1} \right) Y_{L1}-  \frac{9}{4} g_L^{2} Y_{L1}
\end{align*}
\begin{align*}
\beta^{(1)}(Y_{L2}) =&~ Y_{L1} Y_{L1}^{\dagger} Y_{L2}+ \frac{3}{2} Y_{L2} Y_{L2}^{\dagger} Y_{L2}+ 3 {\text{Tr}}\left(Y_{Q1}^{\dagger} Y_{Q1} \right) Y_{L2}+ {\text{Tr}}\left(Y_{L2}^{\dagger} Y_{L2} \right) Y_{L2} -  \frac{3}{4} g_{BL}^{2} Y_{L2}-  \frac{9}{4} g_L^{2} Y_{L2}
\end{align*}
\begin{align*}
\beta^{(1)}(Y_{L3}) =&~ Y_{L1}^{\dagger} Y_{L1} Y_{L3}+ \frac{3}{2} Y_{L3} Y_{L3}^{\dagger} Y_{L3}+ 3 {\text{Tr}}\left(Y_{Q3}^{\dagger} Y_{Q3} \right) Y_{L3}+ {\text{Tr}}\left(Y_{L3}^{\dagger} Y_{L3} \right) Y_{L3}-  \frac{3}{4} g_{BL}^{2} Y_{L3}-  \frac{9}{4} g_R^{2} Y_{L3}
\end{align*}
{\allowdisplaybreaks

\begin{align*}
\beta^{(1)}(\lambda_1) = &~ 32 \lambda_1^{2}+ 16 \lambda_1 \lambda_2+ 16 \lambda_2^{2}+ 8 \alpha_{12}^{2}+ 8 \alpha_{13}^{2}- 9 g_L^{2} \lambda_1- 9 g_R^{2} \lambda_1+ \frac{9}{8} g_L^{4}+ \frac{3}{4} g_L^{2} g_R^{2}+ \frac{9}{8} g_R^{4} \\&+ 12 \lambda_1 {\text{Tr}}\left(Y_{Q2}^{\dagger} Y_{Q2} \right)+ 4 \lambda_1 {\text{Tr}}\left(Y_{L1}^{\dagger} Y_{L1} \right)- 6 {\text{Tr}}\left(Y_{Q2}^{\dagger} Y_{Q2} Y_{Q2}^{\dagger} Y_{Q2} \right)- 2 {\text{Tr}}\left(Y_{L1}^{\dagger} Y_{L1} Y_{L1}^{\dagger} Y_{L1} \right)
\end{align*}
\begin{eqnarray}\label{eq:betalambda2}
\beta^{(1)}(\lambda_2) =&~ 24 \lambda_1 \lambda_2+ 16 \lambda_2^{2}- 4 \alpha_{12}^{2}+ 8 \alpha_{12} \alpha_{13}- 4 \alpha_{13}^{2}- 9 g_L^{2} \lambda_2- 9 g_R^{2} \lambda_2+ \frac{3}{2} g_L^{2} g_R^{2}+ 12 \lambda_2 {\text{Tr}}\left(Y_{Q2}^{\dagger} Y_{Q2} \right)\nonumber \\ 
&+ 4 \lambda_2 {\text{Tr}}\left(Y_{L1}^{\dagger} Y_{L1} \right)+ 3 {\text{Tr}}\left(Y_{Q2}^{\dagger} Y_{Q2} Y_{Q2}^{\dagger} Y_{Q2} \right)
+ {\text{Tr}}\left(Y_{L1}^{\dagger} Y_{L1} Y_{L1}^{\dagger} Y_{L1} \right)
\end{eqnarray}
\begin{align*}
\beta^{(1)}(\lambda_{3}) = &~ 24 \lambda_{3}^{2}+ 8 \lambda_4^{2}+ 8 \alpha_{12}^{2}+ 8 \alpha_{13}^{2}- 3 g_{BL}^{2} \lambda_{3}- 9 g_L^{2} \lambda_{3}+ \frac{3}{8} g_{BL}^{4}+ \frac{3}{4} g_{BL}^{2} g_L^{2}+ \frac{9}{8} g_L^{4}+ 12 \lambda_{3} {\text{Tr}}\left(Y_{Q1}^{\dagger} Y_{Q1} \right) \\&+ 4 \lambda_{3} {\text{Tr}}\left(Y_{L2}^{\dagger} Y_{L2} \right)- 6 {\text{Tr}}\left(Y_{Q1}^{\dagger} Y_{Q1} Y_{Q1}^{\dagger} Y_{Q1} \right)- 2 {\text{Tr}}\left(Y_{L2}^{\dagger} Y_{L2} Y_{L2}^{\dagger} Y_{L2} \right)
\end{align*}
\begin{align*}
\beta^{(1)}(\lambda_4) = &~ 24 \lambda_{3} \lambda_4+ 8 \lambda_4^{2}+ 4 \alpha_{12}^{2}+ 8 \alpha_{12} \alpha_{13}+ 4 \alpha_{13}^{2}- 3 g_{BL}^{2} \lambda_4-  \frac{9}{2} g_L^{2} \lambda_4-  \frac{9}{2} g_R^{2} \lambda_4+ \frac{3}{8} g_{BL}^{4} + 6 \lambda_4 {\text{Tr}}\left(Y_{Q1}^{\dagger} Y_{Q1} \right)\\&+ 6 \lambda_4 {\text{Tr}}\left(Y_{Q3}^{\dagger} Y_{Q3} \right)+ 2 \lambda_4 {\text{Tr}}\left(Y_{L2}^{\dagger} Y_{L2} \right)+ 2 \lambda_4 {\text{Tr}}\left(Y_{L3}^{\dagger} Y_{L3} \right)
\end{align*}
\begin{align*}
\beta^{(1)}(\alpha_{12}) = &~ 12 \alpha_{12} \lambda_1+ 8 \alpha_{12} \lambda_{3}+ 4 \alpha_{12} \lambda_4+ 12 \alpha_{12}^{2}- 8 \alpha_{12} \alpha_{13}+ 8 \alpha_{13} \lambda_1+ 8 \alpha_{13} \lambda_2+ 4 \alpha_{13} \lambda_{3}+ 4 \alpha_{13} \lambda_4 \\&+ 4 \alpha_{13}^{2}-  \frac{3}{2} \alpha_{12} g_{BL}^{2}- 9 \alpha_{12} g_L^{2}-  \frac{9}{2} \alpha_{12} g_R^{2}-\frac{9}{8} g_L^{4}+ 6 \alpha_{12} {\text{Tr}}\left(Y_{Q2}^{\dagger} Y_{Q2} \right)+ 6 \alpha_{12} {\text{Tr}}\left(Y_{Q1}^{\dagger} Y_{Q1} \right) \\&+ 2 \alpha_{12} {\text{Tr}}\left(Y_{L1}^{\dagger} Y_{L1} \right)+ 2 \alpha_{12} {\text{Tr}}\left(Y_{L2}^{\dagger} Y_{L2} \right)
\end{align*}
\begin{align*}
\beta^{(1)}(\alpha_{13}) = &~ 8 \alpha_{12} \lambda_1+ 8 \alpha_{12} \lambda_2+ 4 \alpha_{12} \lambda_{3}+ 4 \alpha_{12} \lambda_4+ 4 \alpha_{12}^{2}- 8 \alpha_{12} \alpha_{13}+ 6 \alpha_{13} \lambda_1+ 8 \alpha_{13} \lambda_{3}+ 4 \alpha_{13} \lambda_4+ 12 \alpha_{13}^{2}\\&-  \frac{9}{2} \alpha_{12} g_L^{2}+ \frac{9}{2} \alpha_{12} g_R^{2}
-  \frac{3}{2} \alpha_{13} g_{BL}^{2}-  \frac{9}{2} \alpha_{13} g_L^{2}- 9 \alpha_{13} g_R^{2}+ \frac{9}{8} g_L^{4}+ 6 \alpha_{12} {\text{Tr}}\left(Y_{Q1}^{\dagger} Y_{Q1} \right)\\&- 6 \alpha_{12} {\text{Tr}}\left(Y_{Q3}^{\dagger} Y_{Q3} \right)+ 2 \alpha_{12} {\text{Tr}}\left(Y_{L2}^{\dagger} Y_{L2} \right)- 2 \alpha_{12} {\text{Tr}}\left(Y_{L3}^{\dagger} Y_{L3} \right)+ 6 \alpha_{13} {\text{Tr}}\left(Y_{Q2}^{\dagger} Y_{Q2} \right)\\&+ 6 \alpha_{13} {\text{Tr}}\left(Y_{Q3}^{\dagger} Y_{Q3} \right)+ 2 \alpha_{13} {\text{Tr}}\left(Y_{L1}^{\dagger} Y_{L1} \right)+ 2 \alpha_{13} {\text{Tr}}\left(Y_{L3}^{\dagger} Y_{L3} \right)- 6 {\text{Tr}}\left(Y_{Q2}^{\dagger} Y_{Q2} Y_{Q3} Y_{Q3}^{\dagger} \right)\\&- 2 {\text{Tr}}\left(Y_{L1}^{\dagger} Y_{L1} Y_{L3} Y_{L3}^{\dagger} \right)
\end{align*}
}

{\allowdisplaybreaks

\begin{align*}
\beta^{(1)}(\mu_1^2) = &~ -  \frac{9}{2} g_L^{2} \mu_1^2-  \frac{9}{2} g_R^{2} \mu_1^2+ 2 \mu_3^{2}+ 20 \lambda_1 \mu_1^2+ 8 \lambda_2 \mu_1^2+ 8 \alpha_{12} \mu_{2}^2+ 8 \alpha_{13} \mu_{2}^2+ 6 \mu_1^2 {\text{Tr}}\left(Y_{Q2}^{\dagger} Y_{Q2} \right) \\ &+ 2 \mu_1^2 {\text{Tr}}\left(Y_{L1}^{\dagger} Y_{L1} \right)
\end{align*}
\begin{align*}
\beta^{(1)}(\mu_{2}^2) = &~-  \frac{3}{2} g_{BL}^{2} \mu_{2}^2-  \frac{9}{2} g_L^{2} \mu_{2}^2+ 4 \mu_3^{2}+ 8 \alpha_{12} \mu_1^2+ 8 \alpha_{13} \mu_1^2+ 12 \lambda_{3} \mu_{2}^2+ 8 \lambda_4 \mu_{2}^2  + 6 \mu_{2}^2 {\text{Tr}}\left(Y_{Q1}^{\dagger} Y_{Q1} \right)\\ &+ 2 \mu_{2}^2 {\text{Tr}}\left(Y_{L2}^{\dagger} Y_{L2} \right)
\end{align*}
}

{\allowdisplaybreaks

\begin{align*}
\beta^{(1)}(\mu_3) = &~-  \frac{3}{2} g_{BL}^{2} \mu_3-  \frac{9}{2} g_L^{2} \mu_3-  \frac{9}{2} g_R^{2} \mu_3+ 4 \lambda_4 \mu_3+ 16 \alpha_{12} \mu_3- 8 \alpha_{13} \mu_3+ 3 \mu_3 {\text{Tr}}\left(Y_{Q2}^{\dagger} Y_{Q2} \right) \\ &+ 3 \mu_3 {\text{Tr}}\left(Y_{Q1}^{\dagger} Y_{Q1} \right)+ 3 \mu_3 {\text{Tr}}\left(Y_{Q3}^{\dagger} Y_{Q3} \right)+ \mu_3 {\text{Tr}}\left(Y_{L1}^{\dagger} Y_{L1} \right)+ \mu_3 {\text{Tr}}\left(Y_{L2}^{\dagger} Y_{L2} \right)+ \mu_3 {\text{Tr}}\left(Y_{L3}^{\dagger} Y_{L3} \right)
\end{align*}
}




\newpage
\printbibliography
\end{document}